\documentclass[referee,astron]{aa}

\usepackage{epsfig}
\usepackage{graphicx}

\usepackage{graphicx}

\begin{document}

\title{Elementary Heating Events - Magnetic Interactions Between
Two Flux Sources. III Energy Considerations}
\titlerunning{Elementary Heating Events}

\author{K. Galsgaard\inst{1} \and C. E. Parnell\inst{2}}
\institute{Niels Bohr Institute, 
Julie Maries Vej 30, 2100 Copenhagen {\O}, Denmark \and School of
Mathematics and Statistics, University of St Andrews, St Andrews, KY16
9SS, Scotland}

\offprints{K. Galsgaard, kg@astro.ku.dk}
\date{Received ; accepted}

\abstract{
The magnetic field plays a crucial role in heating the solar corona -
this has been known for many years - but the exact energy release
mechanism(s) is(are) still unknown. Here, we investigate in detail,
using resistive, non-ideal, MHD models, the process of magnetic energy
release in a situation where two initially independent flux systems
are forced into each other. Work done by the foot point motions goes
into building a current sheet in which magnetic reconnection releases
some of the free magnetic energy leading to magnetic connectivity
changes. The scaling relations of the energy input and output are
determined as functions of the driving velocity and the strength of
fluxes in the independent flux systems. In particular, it is found
that the energy injected into the system is proportional to the
distance travelled. Similarly, the rate of Joule dissipation is
related to the distance travelled. Hence, rapidly driven foot points
lead to bright, intense, but short-lived events, whilst slowly driven
foot points produce weaker, but longer-lived brightenings. Integrated
over the lifetime of the events both would produce the same heating if
all other factors were the same.  A strong overlying field has the
effect of creating compact flux lobes from the sources. These appear
to lead to a more rapid injection of energy, as well as a more rapid
release of energy. Thus, the stronger the overlying field the more
compact and more intense the heating. This means observers need to know 
not only the flux of the magnetic fragments involved in an event, but 
also their rate and direction of movement, as well as the strength 
and orientation of the surrounding
field to be able to predict the energy
dissipated. Furthermore, it is found that rough estimates of the
available energy can be obtained from simple models, starting from
initial potential situations, but that the time scale for the energy
release and, therefore its impact on the coronal plasma, can only be
determined from more detailed investigations of the non-ideal
behaviour of the plasma.
}

\maketitle

\section{Introduction} 
Discussions about the mechanism(s) that maintain the solar corona at
temperatures in excess of a million degrees Kelvin have been going on
since the late fifties.  Much has been learnt, but the exact details
of any mechanism are still not certain. The energy reservoir driving
the heating has, for a long time, been believed to be the turbulent
convection zone below the solar photosphere. Here, the magnetic field
is, to a high degree, frozen into the plasma and turbulent velocity
flows advect the embedded magnetic field.  From the photosphere the
magnetic field extends into the chromosphere and corona.  Buffeting of
the coronal field's `foot points' injects energy that propagates up
along magnetic field lines into the corona. Here, somehow, it is
released contributing to plasma heating, bulk plasma acceleration and
localised particle acceleration.  

Many mechanisms have been investigated and a general division of
models, depending on the timescale of the imposed driver relative to
the Alfv{\'e}n crossing time of the magnetic structure, has been
applied. Boundary motions changing faster than the Alfv{\'e}n travel
time correspond to the initiation of wave packages or trains (AC
heating) that, to a certain degree, propagate along magnetic field
lines and may release their energy through processes such as phase
mixing or resonant absorption 
(e.g., \cite{Heyvaerts+Priest83,Goossens+Ruderman95,Goedbloed79,Ionson78}).
For long time-scale systematic driving periods (DC heating) changes in
the magnetic field structure result in the build up of localised
current sheets. These eventually dissipate through magnetic
reconnection which changes the field line connectivity leading to the
release of the stresses built up in the system
(e.g., \cite{Parker72,Parker88b,Heyvaerts+Priest84,vanBall86,Galsgaard+Nordlund95xc}). 

From small-scale energy release events in the solar atmosphere, such
as ``bright points'' and ``active-region transient brightenings'' a
scenario where the energy release is caused by magnetic flux
cancellation or emergence has been suggested (e.g., 
\cite{Dreher+ea97,Longcope98,Longcope+ea01,Mandrini+ea96,Parnell+ea94,Parnell+ea94b,Parnell+Priest95,Priest+ea94b,Shimizu+ea94}).
Due to the complexities
of the magnetic carpet and the associated overlying magnetic field
structure there is another, possibly more important, mechanism that
has only recently been considered.  In such a situation, the flux
sources are not cancelled or emerged, but are simply advected past each
other resulting in flux connectivity changes as their flux lobes,
which extend into the corona, are forced into each other.
\cite{Longcope98} investigated this scenario for two flux sources
embedded in an overlying magnetic field using the {\it minimum current
corona} approach and found that both the closing and the reopening of
the magnetic flux occurs through separator reconnection. This process
has further been analysed using a numerical approach to solve the
non-ideal time dependent MHD equations
(\cite{Galsgaard+ea00carpet,Parnell+Galsgaard04}).

In these two previous papers (\cite{Galsgaard+ea00carpet} paper I and
\cite{Parnell+Galsgaard04} paper II) different aspects of this type of
interaction were investigated. Two initially independent flux systems
lying in a horizontal uniform field were forced into each other by
imposed boundary motions.  The resulting dynamical interaction,
described in paper I, revealed that there are two types of
reconnection involved.  First, the flux from the independent systems
is connected through separator reconnection and later it is re-opened
through the generally slower separatrix-surface reconnection. The
reconnection rates of these processes were determined in Paper II as
is their dependence on the direction of the overlying magnetic field
with respect to the imposed driver.

In this paper, we investigate the energy dissipation in the same basic
setup with particular attention paid to the effects of the
driving velocity and the strength of the overlying magnetic field.
This enables us to find scaling relations for the energy release in
terms of these parameters. Such scalings are required to provide
predictions of the energy release in similar individual events on the
Sun. We also consider how variations in the driving velocity and field
strength affect the reconnection rate providing further information
necessary for determining the heating capability of such events.  The
structure of the paper is as follows. In Section 2, a brief
description of the model is given. For completeness, Section 3
outlines the general dynamical evolution of the event, as discussed in
detail in Papers I and II. In Section 4, the rates of reconnection and
the connectivity changes in the models are compared. Section 5
discusses issues relating to the energetics of the
experiments. Finally, Section 6, considers the implications of our
findings.

\section{Model}
\begin{figure}
\centering
\epsfxsize=13.5cm
\epsfbox{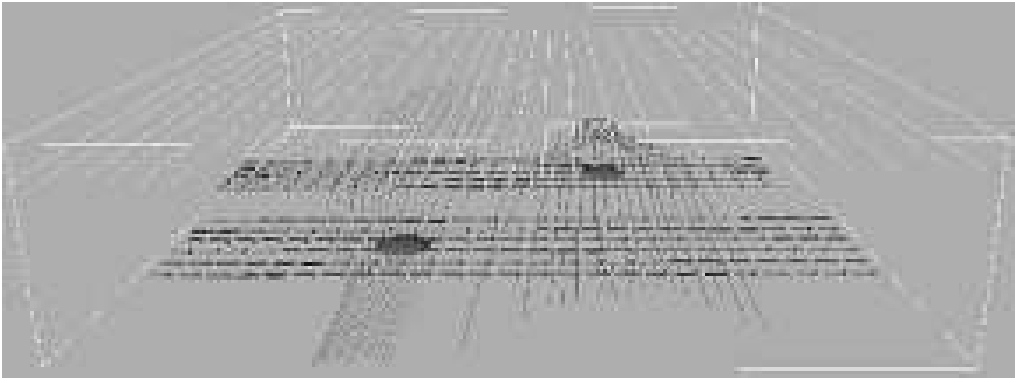}
\caption[]{\label{initial.fig} An example of the initial conditions
      for the experiments.  Magnetic field lines are traced revealing
      the fan separatrix surfaces which outline the independent flux
      lobes from the sources. The shaded regions on the bottom plane
      show the location and flux distribution of the magnetic sources.
      The arrows on the bottom boundary indicate the regions where the
      driving velocity is imposed.
  }
\end{figure}

The setup used here is the same as that used in Papers I and II, where
two localised magnetic sources of equal, but opposite, polarity flux
are situated in a photospheric plane. From these a potential magnetic
field is found numerically for a cubic domain with closed
boundaries. This setup provides a dipole configuration with all the
flux from the positive source connecting to the negative source. In
our domain, which spans [0-1] in the $x$ and $y$ directions, the two
sources are located at $(x,y)=(\frac{1}{3},\frac{1}{3})$ and
$(x,y)=(\frac{2}{3},\frac{2}{3})$ on the $z=0$ boundary of the
domain. The flux within the sources follows a $\frac{1}{2}(1+\cos(\pi
r/R))$ distribution, where $r$ is the radius from the source centre
and $R$ the maximum radius of the sources (in these experiments $R$ is
0.065 in units of the box). To break the connectivity of the sources a
constant magnetic field, $B_y$, of sufficient strength and the correct
sign, is added in the $y$ direction, thus rendering the two flux
sources magnetically independent with their associated flux lobes
running parallel to each other in the $y$ direction. Two 3D magnetic
null points are created in the $z=0$ plane and are orientated in such
a way that their spine axes lie in the $z=0$ plane, connecting each
null with a single source, whilst their fan planes divide space into
independent flux regions. The initial topology in the experiments
(shown in Fig. \ref{initial.fig}) has three independent flux regions -
two connected to the sources and one containing the overlying field.

The imposed driving in these experiments has the same form as that
used in papers I and II, namely a piecewise uniform advection
of the flux sources in the $x$ direction such that the independent
flux regions are driven past each other, see
Fig. \ref{initial.fig}. The velocity is imposed in two narrow regions
on the $z=0$ boundary with sufficient width such that the sources are
advected without changing shape.  We are interested, here, in the
dynamical evolution of the magnetic field and, therefore, we ignore the
complicated structure of the solar atmosphere and instead consider,
for simplicity, an isothermal, constant density atmosphere.

For numerical reasons, the closed box, which has no flux passing
through the horizontal walls and is used to derive the initial
potential magnetic field, is replaced by a domain with boundaries that
are periodic in the horizontal directions (allowing flux through these
walls) during the time dependent evolution of the system. We could
have derived the initial potential magnetic field of such a 2D
periodic configuration, but this has the disadvantage that the initial
connectivity of the sources becomes much more complicated as they
would be allowed to connect to other sources through all the periodic
boundaries, as well as inside the domain. The disadvantage with
having a change in boundary conditions between the initial and time
dependent field is that the periodic (horizontal) sides of the box
create a narrow layer where initial spurious current concentrations
form and the condition $\nabla\cdot{\bf B} = 0$ is not fulfilled. The
currents initiate waves that propagate through the domain with
time. Their amplitudes are insignificant compared to the dynamical
response of the plasma to the advection of the two flux sources and
have no implications on the evolution of the magnetic field. The
regions where $\nabla\cdot{\bf B} \neq 0$ remain fixed and do not
propagate through the domain so they cause no problems for the
evolution of the magnetic field.

Two sets of experiments are investigated to better understand the
energy release and reconnection processes as the sources are forced to
pass each other.  One set considers the implications of different
driving speeds of the sources whilst the second allows the effects of
varying the relative strengths of the flux sources with respect to the
overlying field to be investigated.  For the first set, the magnetic
configuration is kept constant, but the driving velocity in the $z=0$
boundary is varied. In the second case, the strength of the overlying
magnetic field in the $y$ direction is varied.  To limit the number of
experiments and to allow easy comparisons, the range of overlying
field strengths are chosen such that initially the two sources are
always totally unconnected.  The minimum value of $B_y$ used is such
that the flux lobes of the two sources are nearly touching in the
$z=0$ plane. Similarly, the maximum value is chosen such that
sufficient numerical resolution of the flux domains is maintained.
Changing the $B_y$ component is equivalent to changing the length
scale of the independent flux regions: as $B_y$ is increased the
magnetic flux per unit area increases.

Table (\ref{cases.tab}) shows the characteristic parameters of the
experiments that are conducted, listing the imposed peak driving
velocity $v_d$, the duration of driving $t_d$, the strength of $B_y$
and the numerical resolution (all parameters are given in units of the
code).
\begin{table}[ht]
\begin{center}
\caption[]{\label{cases.tab} The characteristic parameters of the 8
           experiments, giving the name, driving velocity and
           duration, the strength of the constant overlying magnetic
           field and the numerical resolution. Note, experiments $D2$
           and $H2$ are the same.}
\begin{tabular}{c l c c c} \hline
Case & $v_d$ & $t_d$ & $B_y$ & resolution \\ \hline
$D1$ & 0.0125 & 44 & 0.1 & 128x128x65\\
$D2$ & 0.025 & 22 & 0.1 & 128x128x65\\
$D3$ & 0.05 & 11 & 0.1 & 128x128x65\\ \hline
$D4$ & 0.025 & - & 0.1 & 128x128x65\\ \hline
$D5$ & 0.025 & 22 & 0.1 & 256x256x129\\ \hline
$H1$ & 0.025 & 22 & 0.076 & 128x128x65\\
$H2$ & 0.025 & 22 & 0.1 & 128x128x65\\
$H3$ & 0.025 & 22 & 0.2 & 128x128x65\\ \hline
$H4$ & 0.025 & 22 & 0.2 & 256x256x129\\ \hline
\end{tabular}
\end{center}
\end{table}

Together these experiments are designed to increase our insight into
the dynamical interaction of magnetic flux systems and provide a basis
for predicting energy release rates in similar events observed in the
solar atmosphere.

\section{Global Behaviour}
For completeness a short description of the dynamical evolution of the
magnetic field is given here; a more detail account of the evolution
can be found in papers I and II.  The dynamical evolution is followed
by solving the time dependent, non-ideal, MHD equations numerically in
a 3D Cartesian domain (for more details about the numerical approach
see paper I and II and references their in).

The opposite polarity magnetic sources are advected by the imposed
boundary flow and their flux lobes move towards each other. When the
flux lobe from one source comes up against the other source, the
flux lobe is lifted up over the moving source.  This action leads to
an interlocking of the flux lobes from the sources. Further advection
of the sources creates forces on the intertwined flux lobes and a
strong current builds up along the interface between them.  This
current is located along the separator line connecting the two
magnetic null points which, due to the continued advection, collapses
into a current sheet.  When the current density becomes strong enough
reconnection starts within this sheet rapidly changing the field line
connectivity and allowing the two sources to connect. This process
continues until the sources are well past the point of closest
approach and a large fraction of the initially open flux is closed.

As the sources are advected even further apart an interface between
the connected flux and the ambient open field creates a dome shaped
separatrix surface upon which strong, some what irregular, current
concentrations form. This current is a consequence of the rapid
tangential change in orientation of the field lines across this
surface and is associated with the reopening of the magnetic field
connecting the two sources. When the driving is stopped, in general,
only a fraction of the connected flux has re-opened and the opening
rate decreases rapidly, requiring a long time to fully disconnect the
two sources.

\section{\label{drive_connect.sec}Connectivity Changes and Reconnection Rates}
In this paper, we are interested in studying the energetics of simple
magnetic interactions. In order to equate changes seen in the
energetics of the experiments to physical changes in the system, we
first discuss briefly the connectivity changes and rates of
reconnection that occur.  The results for the varying driving speed
experiments are the same as those given in paper II, however, the
varying overlying field strength experiments are all new.

\subsection{Changes in Connectivity}
\begin{figure}
\centering
\scalebox{0.5}{\includegraphics*{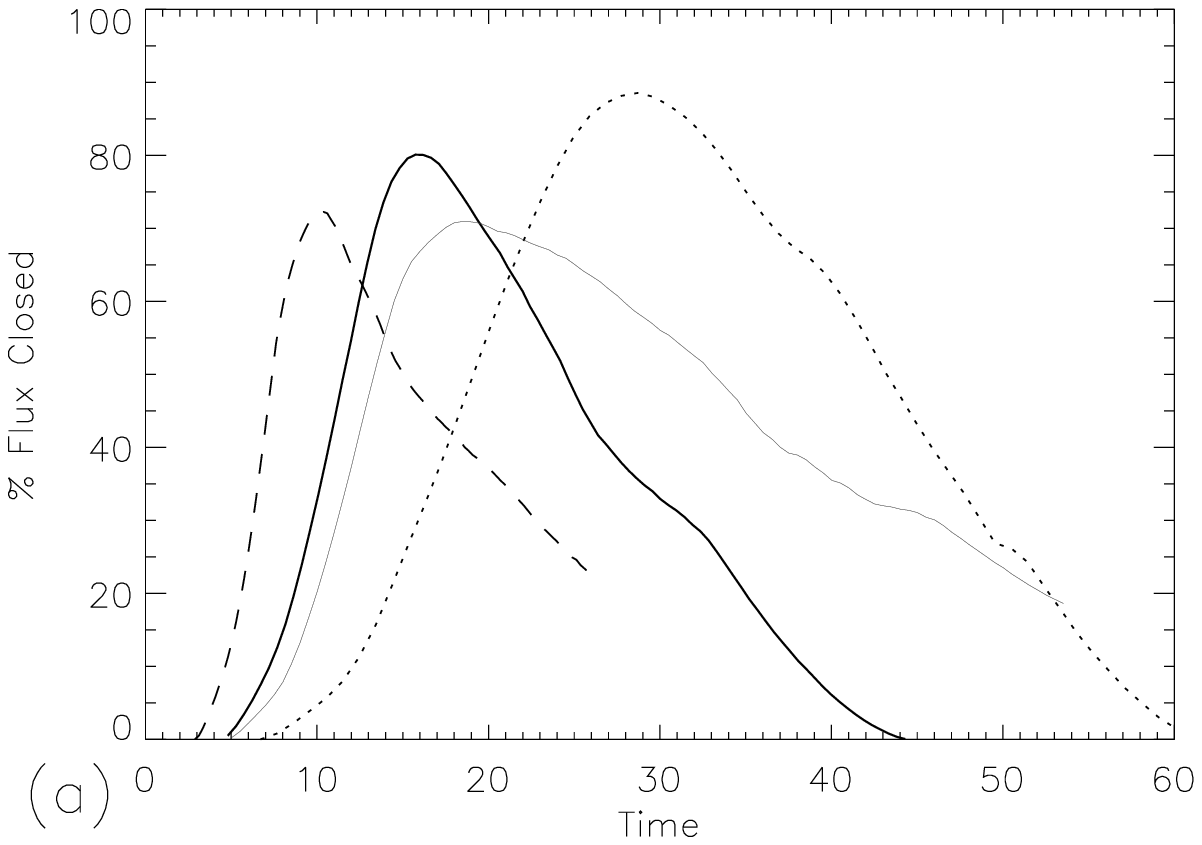}}
\scalebox{0.5}{\includegraphics*{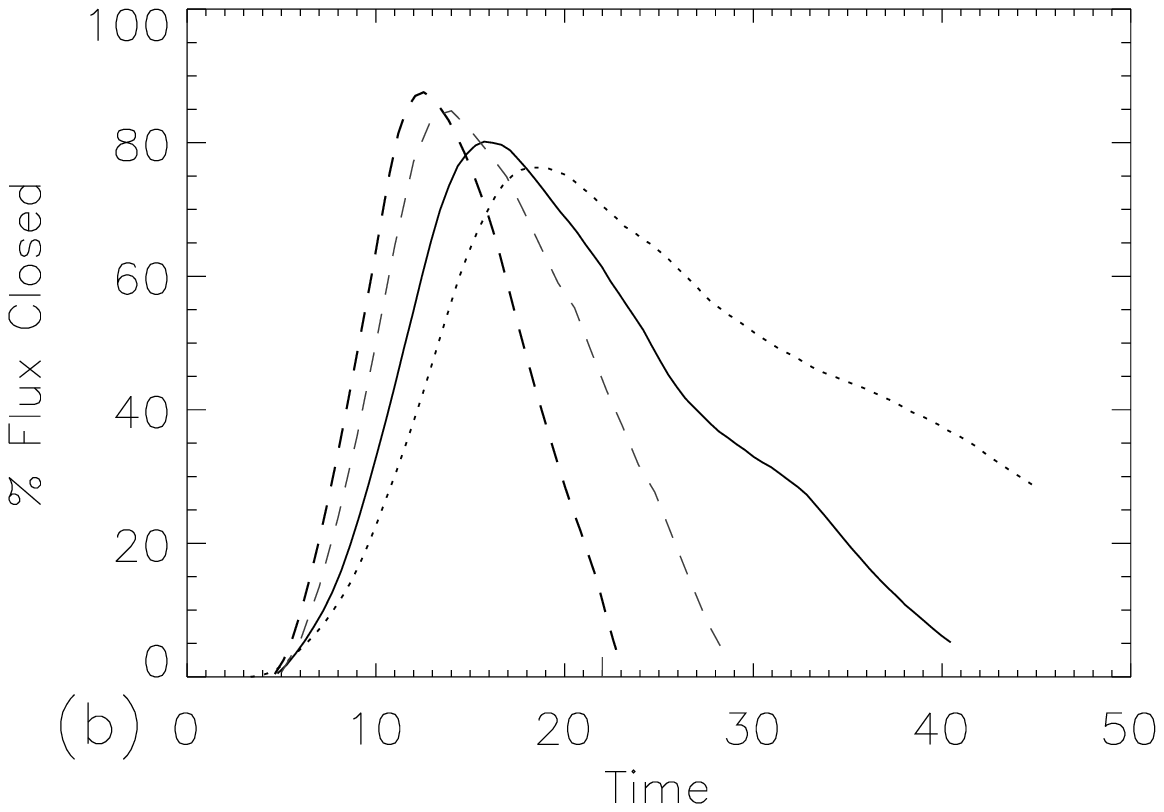}}
\scalebox{0.5}{\includegraphics*{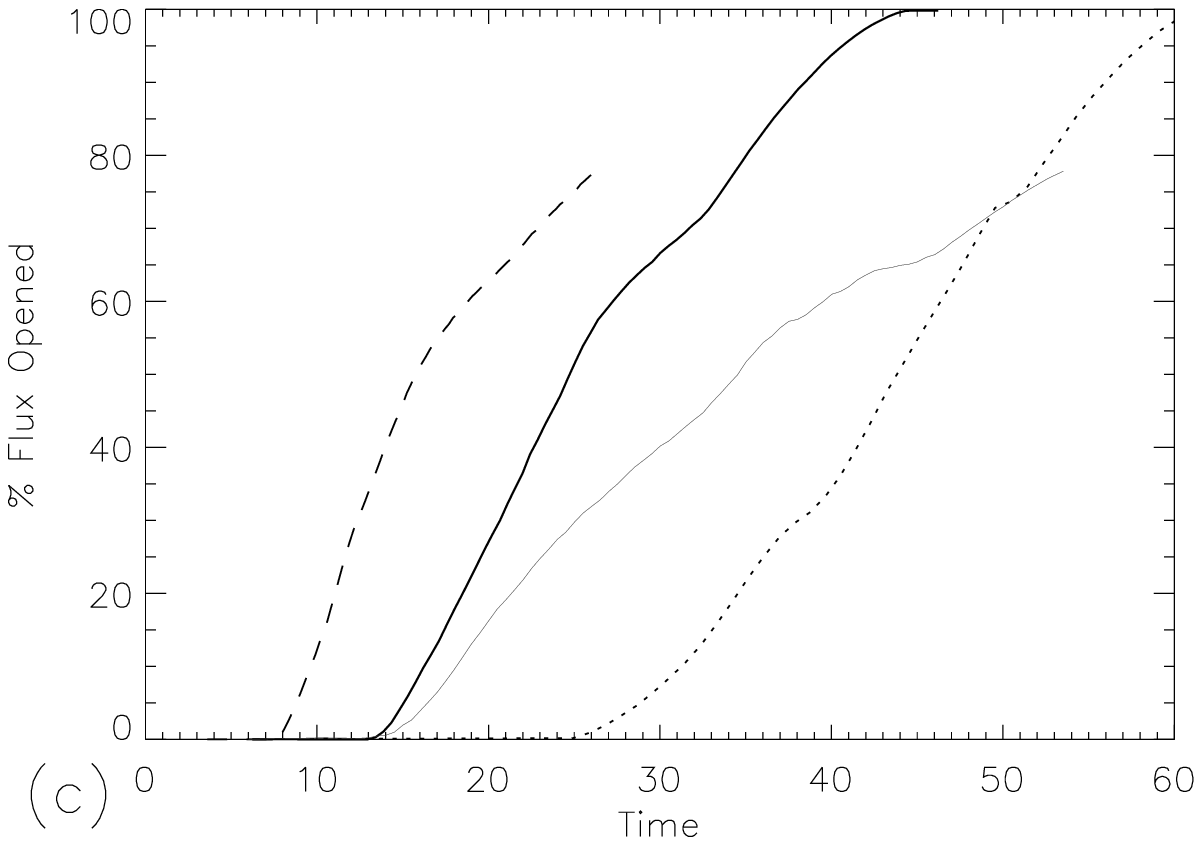}}
\scalebox{0.5}{\includegraphics*{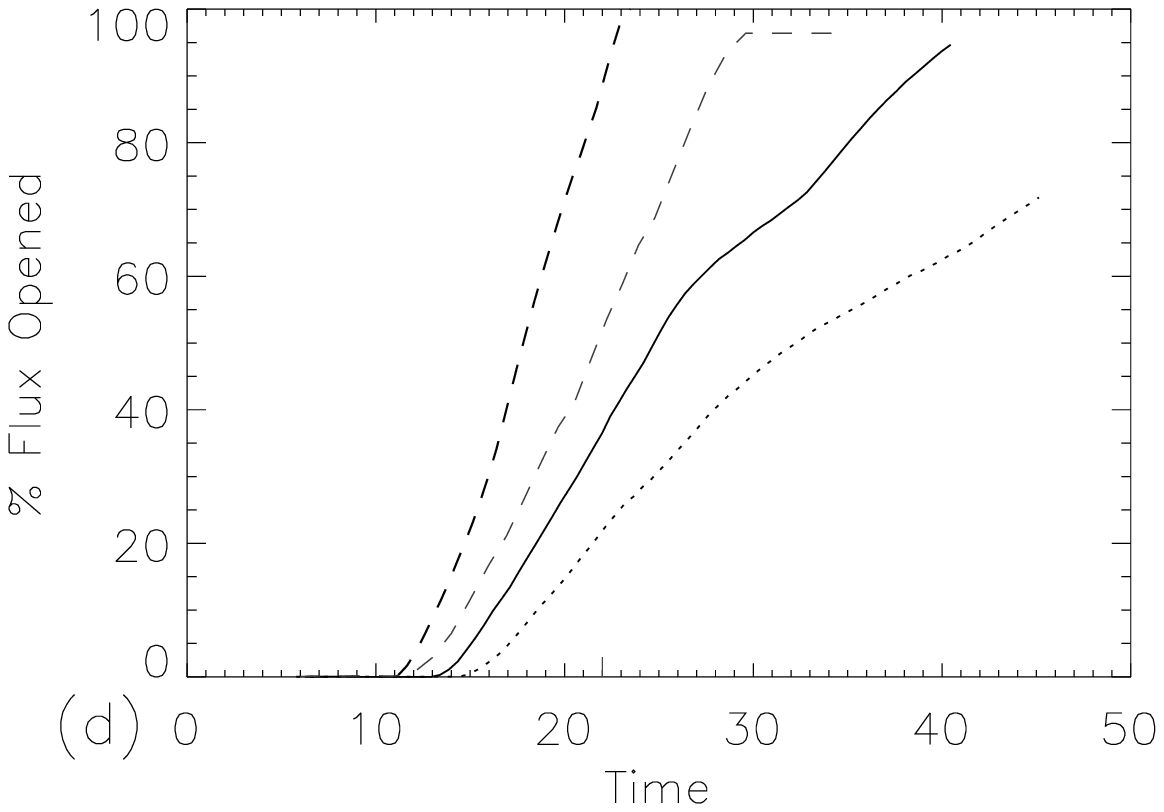}}
\scalebox{0.65}{\rotatebox{90}{\includegraphics*{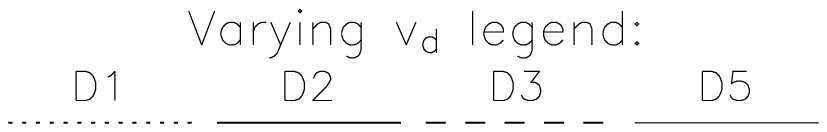}}}
\scalebox{0.65}{\rotatebox{90}{\includegraphics*{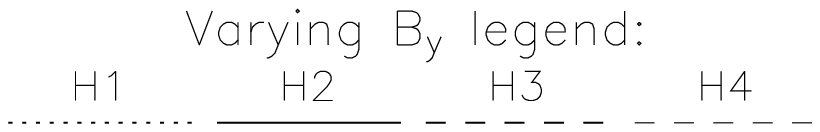}}}
\caption[]{\label{con_rec_driv_height.fig} The percentage of (a \& b)
   closed and (c \& d) re-opened flux versus time. The graphs on the
   left refer to the varying driving speed experiments and those on
   the right to the varying overlying field strength experiments.}
\end{figure}

As in Paper II, the change in field line connectivity between the two
sources is followed for each experiment. The graphs in
Fig. \ref{con_rec_driv_height.fig} show the temporal evolution of the
percentage of closed flux (top row) or percentage of re-opened flux
(second row).  The two left-hand graphs show the curves for the
varying driving speed experiments whilst the right-hand ones show
those with varying overlying field strength. In all experiments, the
time is in numerical time units. The percentage of closed or re-opened
flux for each source is found by identifying the endpoints of field
lines (thin flux tubes) in exactly the same way as those in paper
II. More than 12 500 flux tubes are tracked from each source at each
time step with a maximum of 0.03 \% of the total source flux in any
one flux tube.

Although the two sources in each experiments are initially completely
unconnected they all pass through a phase where they become increasing
connected.  Just before the amount of closed flux reaches its peaks,
the first re-opened flux is generated.  There are a number of points
that can be noted from these graphs:

\begin{itemize}
\item The time of onset of closing the flux is dependent on the speed
of driving (e.g., $t=8$, 5, 3 for $v_d=0.0125$, 0.025 and 0.05), but
appears to be independent of the strength of the overlying field
(e.g., $t=5$ for all $B_y$). Basically, the separator reconnection
process which creates the closed field starts when sufficient stresses
(currents) have developed between the two entwined flux lobes and,
hence, is dependent on how far the flux lobes have moved. The
dependence on the driving speed is therefore not surprising. Note,
that the more slowly driven sources start to connect after a shorter
distance has been travelled than the more quickly driven sources. This
is due to the travel time of information in the domain which can
respond relatively faster the more slowly it is driven.  In the
experiments with varying overlying field strength the sources start to
connect at about the same time. At first sight this seems strange
since the overlying field has the affect of decreasing the size of the
flux lobes as its strength increases. Thus, for the same advection
distance, the strongest overlying field experiments produces the least
entwined flux lobes. However, this effect is counteracted by the fact
that the flux density within the lobes is greater with higher $B_y$,
leading to similar currents being generated and thus reconnection
starting at about the same time in each experiment.

\item The size and time of the peak closed flux is dependent on both
the driving speed and the strength of the overlying field. The fastest
driver and strongest overlying field both lead to the earliest peaks
in closed flux. This is because in both cases they give rise to rapid
reconnection since the fast driving and strong field maintain the
stresses in the system. Note, however that the faster the driver the
lower the peak in closed flux suggesting that the more rapid the
reconnection the less complete it is. On the other hand the
reconnection appears to be more complete for stronger $B_y$ leading to
a greater peak in closed flux. The completeness of the reconnection
process is determined by the relative speed of the driver to the
Alfv{\'e}n travel time in the system. In the fast driver case there is
little time for the system to respond, however, in the strong $B_y$
case the Alfv{\'e}n speed is greater and so the system has more time to
respond.

\item The start of the re-opening process (separatrix-surface
reconnection) is affected by both $v_d$ and $B_y$. Again the more
rapid the driver the earlier in time the re-opening starts, $t=25$, 14
and 8, respectively, for $v_d=0.0125$, 0.025 and 0.05. However, by
comparing the advection distance instead it is the slowest driven case
that starts to reopen at the shortest driven distance. It therefore
also has the shortest advection distance between the onsets of the two
classes of reconnection.  The onset of reopening also depends on the
overlying field strength, with the strongest $B_y$ seeing the first
reopened flux. Thus, the system with the shortest Alfv{\'e}n travel
time (largest value of $B_y$) responds the fastest.
\end{itemize}

All the above points indicate that, of course, reconnection in a
dynamical MHD situation is not rapid enough to process all the flux as
soon as it is advected into the current sheet, as would happen through
equi-potential evolution.

In the classical 2D reconnection scenarios
(\cite{Sweet58,Parker57,Petschek64}) the reconnection rate scales
inversely as some function of the magnetic Reynolds number. The
reconnection rate can therefore be changed by changing any of the
three parameters defining the Reynolds number.  For instance,
increasing the coronal field strength leads to a decrease in the
length of the current sheet and, hence, an increase in the
reconnection rate (assuming this is the only parameter
changing). However, from Fig. \ref{con_rec_driv_height.fig}, it is
seen that in the 3D numerical experiments the situation is more
complicated than this. An increase in overlying magnetic field
strength not only leads to a decrease in width of the current sheet
(Note, in three-dimensions, a current sheet typically has dimensions
thickness $<<$ width $<<$ length, as seen in Fig. 2
(\cite{Parnell+Galsgaard04}) and so for the local magnetic Reynolds
number what is important is the width of the current sheet), but also
results in an increase in the local Alfv{\'e}n speed in the vicinity
of the reconnection site. Since two factors have now changed the
resulting change in the rate of reconnection is not clear and a priori
cannot be predicted. In fact, it seems that an increase in overlying
field strength leads to an increase in reconnection rate. Further,
increasing the numerical resolution, effectively decreases $\eta$ and
subsequently the reconnection speed. Due to the multiple effects of
changing a single parameter in the experiments it is not simple to
predict what would happen as various parameters are changed. Instead,
by obtaining the magnetic Reynolds numbers for each experiment it
is possible to show that the behaviour follows the classical trend,
with a decreasing reconnection rate for increasing $R_m$.
   
The effect of higher resolution leading to delayed and slightly slower
reconnection is again seen by comparing $D5$, a high resolution run
with $D2$ its lower resolution counterpart. Otherwise these two runs
are essentially the same.

\subsection {Reconnection Rates}
\begin{figure}
\centering
\scalebox{0.5}{\includegraphics*{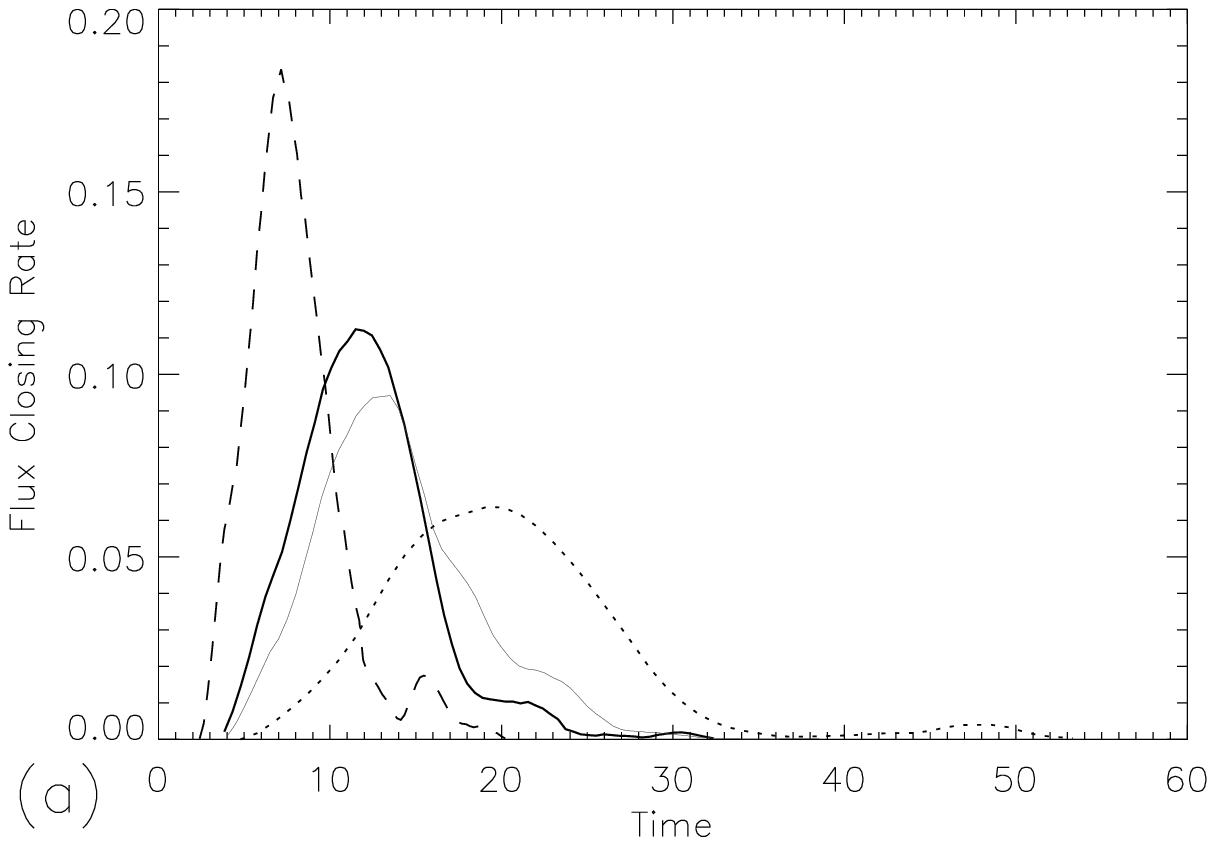}}
\scalebox{0.5}{\includegraphics*{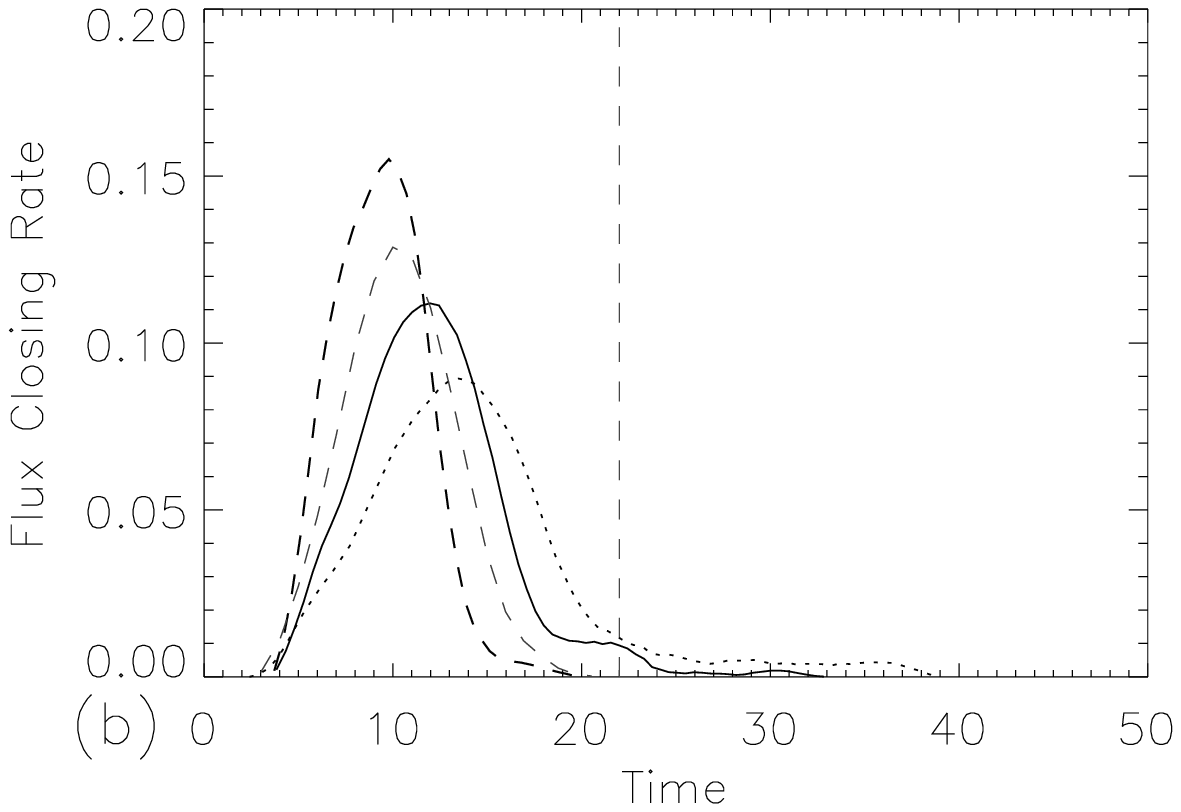}}
\scalebox{0.5}{\includegraphics*{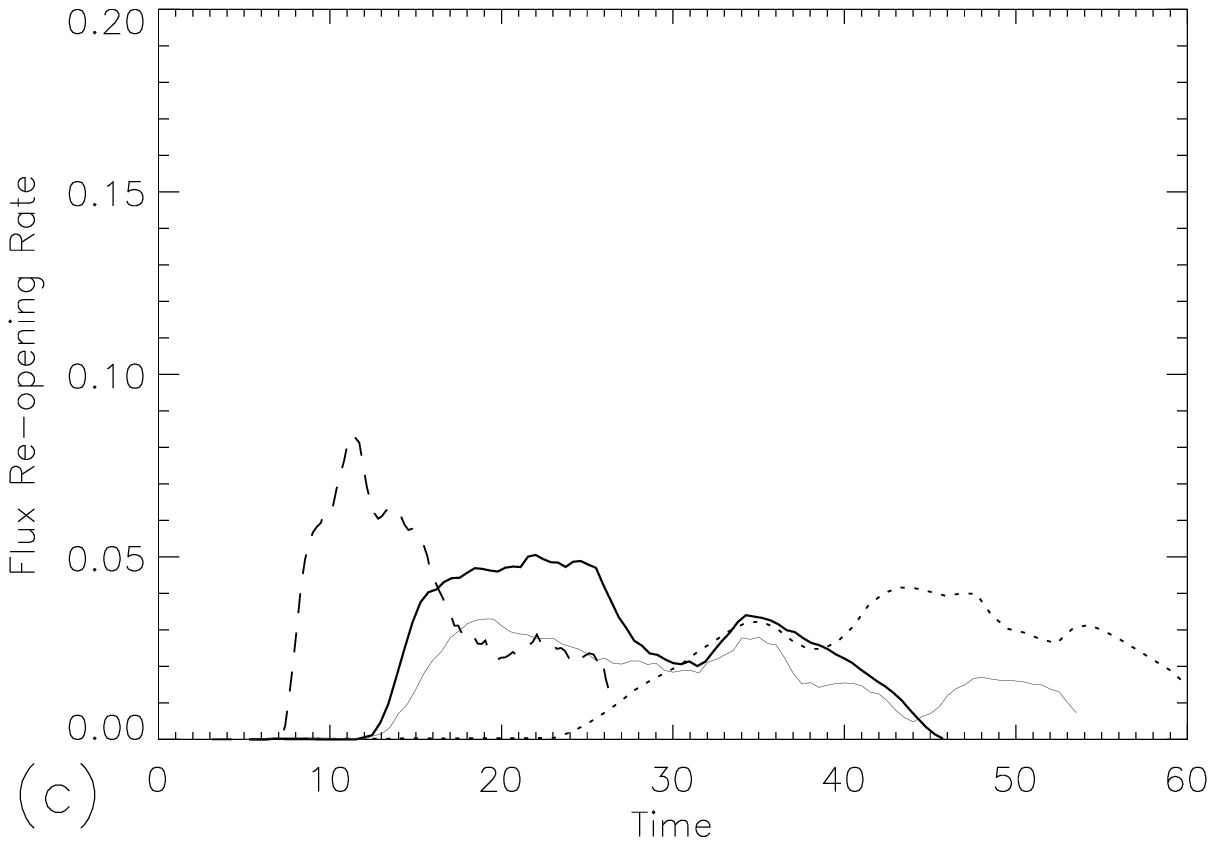}}
\scalebox{0.5}{\includegraphics*{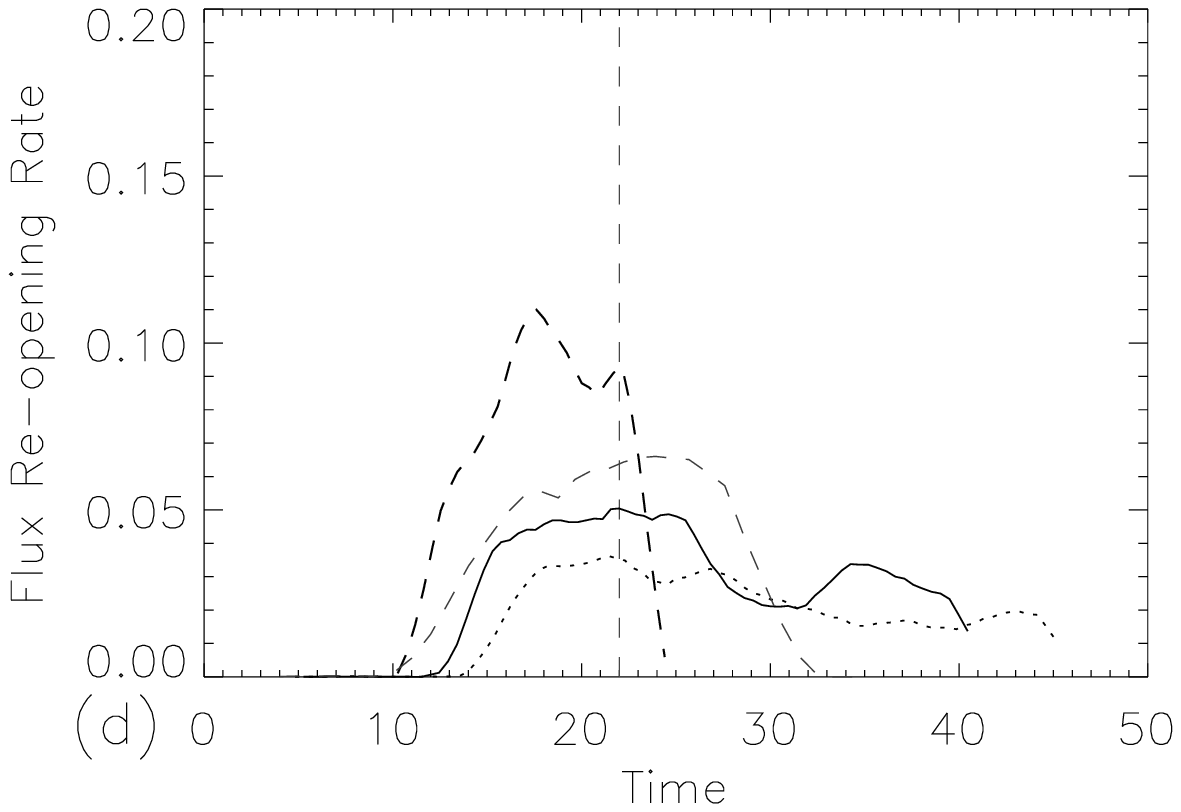}}
\scalebox{0.65}{\rotatebox{90}{\includegraphics*{PS/legend_drive_2.eps}}}
\scalebox{0.65}{\rotatebox{90}{\includegraphics*{PS/legend_height.eps}}}
\caption[]{\label{rec_driv_height.fig} The reconnection rate of (a \&
   b) closing and (c \& d) opening the field. The left-hand graphs
   refer to the varying driving speed experiments and those on the
   right to the varying overlying field strength experiments.}
\end{figure}

The two rows of graphs in Fig. \ref{rec_driv_height.fig} show the
temporal evolution of the separator (closing) and separatrix-surface
(opening) reconnection rates, respectively. With, as before, the
varying driving speed results on the left and the varying overlying
field results on the right.  The rates of reconnection have been
determined in exactly the same way as those in paper II, i.e., from
differentiation of the flux closing and re-opening curves. From these
graphs it is clear that the peak rate of closing or opening the field
is dependent on the speed of the drivers. From paper II the following
formulae were determined relating driving speed ($v_d$) and overlying
field angle ($\theta$ - here $\theta=0$) to peak closing ($R_{cmx}$)
or re-opening ($R_{omx}$) reconnection rates:
$$R_{cmx}=\frac{0.44v_d\cos\theta}{v_d\cos\theta+0.08},\hspace{0.5cm}
{\rm and}\hspace{0.5cm}R_{omx}=
\frac{0.19v_d\cos\theta}{v_d\cos\theta+0.06},$$ where in these
formulae the driving speed and peak rates of reconnection are
normalised with respect to the peak Alfv{\'e}n velocity in the
sources. Thus, the peak rates of separator reconnection are about
twice the peak rates of separatrix-surface reconnection. Moreover, it
was found that, at the typical observed driving speed of fragments in
the photosphere (i.e., a hundredth of the Alfv{\'e}n speed) the peak
rate of separator reconnection was 58\% of the instantaneous
reconnection rate (determined from the evolution of an equi-potential
model) and the separatrix-surface reconnection rate was 29\%. Thus, both the separator and separatrix-surface reconnection
rates are fast.

The reconnection rates also vary for varying overlying field
strength. There are two effects that are at work in these
experiments. Varying the overlying field leads to (i) changes in the
size of the flux lobes and, therefore, changes in the flux per unit
area (a real physical effect), and (ii) changing the size of the flux
lobes has the knock on effect of changing the resolution across the
lobe and, hence, changing the resistivity of the experiment (a
numerical artifact).  Separating these two effects is not straight
forward and so determining exactly how the rate of reconnection scales
with $B_y$ is not clear. Neither is the significance of the effect.

\section{Energetics}
Here, we investigate the energetics of the different scenarios given
in Table (\ref{cases.tab}). In particular, we are interested in the
balance between the energy input, the energy growth, and its release
through viscous and resistive processes. 

\subsection{Effects of the Driving Velocity}
First, we discuss the impact of the driving velocity on the energetics
of the system. To make comparisons between the experiments easier, the
time scales of the experiments are scaled to fit that of experiment
$D2$ such that the sources cover the same distance in the same time
(i.e., one could consider that they have been plotted against driving
distance rather than time). This scaling is only applied to the
various parameters up until the driver is switched off, after which
the parameters are all plotted against their true numerical time.
Clearly, where the times are scaled the time dependent variables, such
as the Poynting flux and the dissipation rates, must also be scaled
accordingly, but no changes need be made to the volume energy
parameters.

\subsubsection{Poynting Flux}
\label{Pf_driv.sec}
\begin{figure}
\centering
\epsfxsize=7.cm
\epsfbox{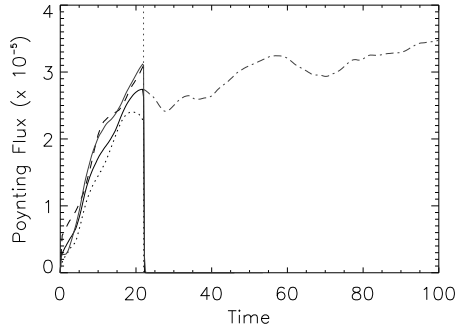}
\caption[]{\label{pf_driv.fig} The scaled Poynting flux for the
   experiments, $D1$-$D5$, as a function of scaled time. The imposed
   driving velocities are 0.0125 (dotted line {\it - $D1$}), 0.025
   (thick solid line {\it - $D2$}, thin solid line {\it $D5$} and
   dot-dashed line {\it $D4$}) and 0.05 (dashed line {\it -
   $D3$}). The vertical dashed line is the time at which all
   drivers are switched off except for $D4$ which is driven
   continuously. $D5$ has higher numerical resolution than all the
   other experiments.}
\end{figure}
Fig. \ref{pf_driv.fig} shows the Poynting flux for the five
experiments, $D1$-$D5$, which are all scaled to the same advection
distance as that used in experiment $D2$ for comparison. This requires
multiplication by a factor of 2 and 0.5, respectively, to the actual
Poynting fluxes of experiments $D1$ and $D3$.  Clearly, all the
experiments have very similar scaled Poynting fluxes which show an
almost linear, growth in time up until the switching off of the
drivers. The undulations in the curves simply correspond to the
Alfv{\'e}n travel time across the periodic domain. It is only towards
the end of the driving period that a turn over in the Poynting flux is
found in the slowest driven experiment $D1$ and also possibly in
$D2$. There is little difference between $D2$ and $D5$, which have the
same driving velocity, but a different numerical resolution.

The three curves {\it $D1$-$D3$} represent changes in driving velocity
with a factor of 2 between each one. So why, when scaled to the same
time frame, do they show almost the same linear growth with time? Also
why are there offsets in the initial Poynting flux values? These
questions can be answered by looking at the expressions for the
induction equation and the Poynting flux.

Consider the ideal induction equation with velocity simply in the $x$
direction, ${\bf u}=(u_x,0,0)$. The resulting changes in the magnetic
field, ${\bf B}= (B_x,B_y,B_z)$, on the driving boundary are given by
\begin{eqnarray}
\label{B_induction.eq}
{{\partial {\bf B}} \over {\partial t}} = \left (
{{\partial } \over {\partial y}} (u_x B_y) 
+{{\partial } \over {\partial z}} (u_x B_z),
-{{\partial } \over {\partial x}} (u_x B_y),
-{{\partial } \over {\partial x}} (u_x B_z) \right ).
\end{eqnarray}
Assuming that (i) the magnetic field is directed into the domain, (ii)
$u_x$ is in the positive $x$ direction (iii) both the magnitudes of
${\bf B}$ and ${\bf u}$ decrease with height (i.e., increasing $z$)
and (iv) both ${\bf B}$ and ${\bf u}$ have only a weak $y$ dependence,
then $B_x$ will have a negative growth in height above the driving
boundary surface.  At the same time both $B_y$ and $B_z$ are advected
with the flow, maintaining their spatial structure.  This indicates
that the magnetic field lines that thread the driving boundary (i.e.,
the magnetic field lines from the sources) lag behind their foot points
as they are driven across the box resulting in a change in the angle
of the magnetic field in the $xz$ plane as time progresses.
Furthermore, assuming that (i) the driving velocity is slow relative to
the Alfv{\'e}n speed and (ii) the time since the driving started is short
in comparison to the Alfv{\'e}n time of the advected field lines, then
the ratio of the driving velocity to the Alfv{\'e}n velocity
determines the maximum angle of the advected magnetic field lines to
the surface normal, $\phi = \mbox{arctan}(v_d/v_A)$. Now the Poynting
flux injected by the driving is
\begin{eqnarray}
\label{Pf_global.eq}
P_{f} = {1 \over \mu} \int ({\bf u} \times {\bf B}) \times {\bf B} 
\cdot \bf{dS},
\end{eqnarray}
where ${\bf u}$ is the velocity on the boundary surface, $S$, and
${\bf B}$ is the magnetic field.  Here, $S$ is the constant $z=0$
surface and, as indicated above, the velocity is in the $x$-direction
only, thus Eq. (\ref{Pf_global.eq}) reduces to
\begin{eqnarray}
\label{Pf_plane.eq}
P_{f} = {1 \over \mu} \int \int u_x B_x B_z dx dy.
\end{eqnarray}
Hence, the Poynting flux is fixed at a constant level proportional to $\phi$. 

In the alternative case, where the driving velocity is also slow, but
the time since the start of the driving is long in comparison to the
Alfv{\'e}n crossing times of the advected field lines, the angle
$\phi$ grows linearly with time as the foot points are constantly
advected in a systematic direction and the Poynting flux grows
linearly with time (as already pointed out by
\cite{Parker87a,Galsgaard+Nordlund95xc}).

In the present situation, neither of these two extremes fully applies,
however, the near-linear growth of the Poynting flux in
Fig. \ref{pf_driv.fig} suggests that the situation here is closest to
the latter case.  The reason for this may lie in the fact that the two
flux lobes from the sources are soon forced into each other
effectively fixing the `free' ends of the field lines thus resulting
in short field lines. Or it may be because of the rapid change in
magnetic field strength as one moves away from the sources resulting in
a decrease in the Alfv{\'e}n velocity and hence, a slowing down of the
propagation speed of information.

By assuming that the $x$ component of the induction equation is only
weakly dependent on $y$, the value of $B_x$ can be approximated by,
\begin{eqnarray}
\label{bx_t.eq}
B_x (t_d) = {\partial \over {\partial z}} (u_x B_z) t_d \approx 
{{u_x B_z t_d }\over L},
\end{eqnarray}
where $L$ is the distance to the (apparent) end points of the field
lines and $t_d$ is the time since the driving started. Substituting
this into Eq. (\ref{Pf_plane.eq}) leads to the following expression
for the average Poynting flux input through $S$,
\begin{eqnarray}
\label{pf.eq}
P_{f} \approx {1 \over \mu} \int\int{{B_z^2 u_x^2 t_d }\over L} dx dy.
\end{eqnarray}

Assuming that $L$ and $B_z$ are close to constant in all the
experiments, implies that changes in driving velocity and driving time
lead to a simple scaling of the different cases. Namely, a doubling of
the velocity increases the Poynting flux by a factor of 2 at the same
shear distance.  Hence, the re-scaling of the experiments for the
graphs produces almost identical values for the experiments.  The turn
over of the Poynting flux in $D1$ indicates where the above
approximation breaks down and a new regime starts.

The differences in Poynting flux seen in the initial phases of the
experiments are effectively maintained throughout the driving
period. Clearly, at the start, the Poynting flux is zero in every
experiment, but when the driving is switched on there is a
discontinuous jump.  The level of the jump depends on the ratio of the
driving and Alfv{\'e}n velocities. This ratio is proportional to the
ratio of $B_x$ and $B_z$, which shows that $B_x$ is proportional to
the driving velocity in this regime. Substituting this into
Eq.(\ref{Pf_plane.eq}) one gets,
\begin{eqnarray}
\label{pf_wave.eq}
P_{f} \approx {{B_z^2 u_x^2}\over {\mu u_A}}.
\end{eqnarray}
where $u_A$ is the Alfv{\'e}n velocity. Doubling the driving
velocity quadruples the Poynting flux which, with re-scaling, results
in a doubling of the initial jump in Poynting flux, as seen in
Fig. \ref{pf_driv.fig}.

In Fig. \ref{pf_driv.fig}, the continuously driven experiment $D4$ is
clearly identifiably as the dot-dashed line that extends from the
solid line of $D1$. Here, the initial linear growth is replaced with a
more constant energy input after t = 22 with some fluctuations in
time. This implies that even after the flux sources have passed each
other a significant amount of energy continues to be injected through
the driving process. This has similarities to the results that
\cite{Galsgaard+Nordlund95xc} found in their experiments investigating
flux braiding.  In their scenario, an initial straight magnetic field
was braided by a sequence of incompressible shear motions and after a
few shear events a similar fluctuating input level, as seen here in
$D4$, was obtained. Obviously, though here in $D4$, to accommodate
such a sustained driving period, the sources are driven out of the box
and back in the opposite side (since the side boundaries are
periodic). Thus, interpretation of the results is not straight forward.

\subsubsection{Magnetic and Kinetic Energy}
\begin{figure}
\epsfxsize=7.cm
\epsfbox{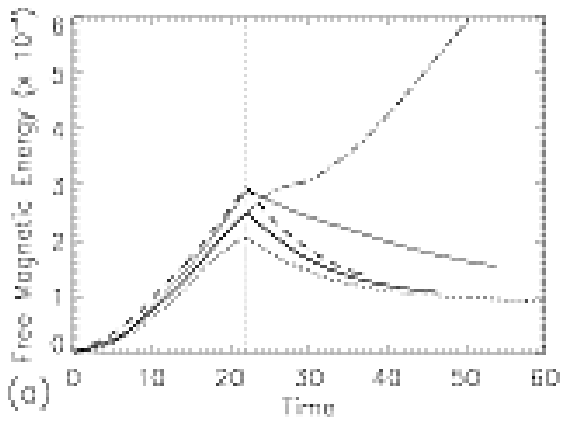}
\epsfxsize=7.cm
\epsfxsize=7.cm
\epsfbox{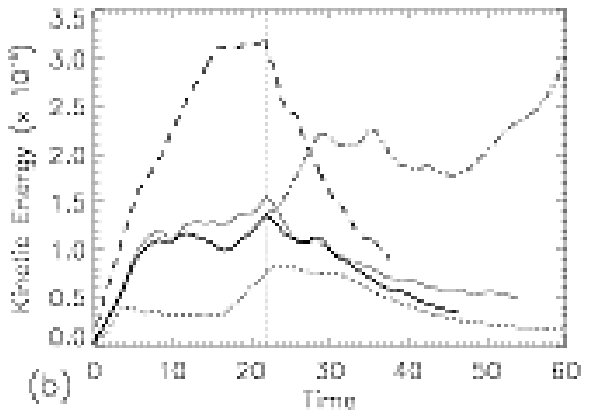}
\caption[]{\label{mag_driv.fig} (a) The `free' magnetic energy
     (magnetic energy minus the potential energy) and (b) the kinetic
     energy for $D1$-$D5$ as a function of scaled time for t$<$22 and
     as a function of the correct time for t$>$22. The velocities are
     0.0125 (dotted line {\it - $D1$}), 0.025 (thick solid line {\it -
     $D2$}, thin solid line {\it $D5$} and dot-dashed line {\it $D4$})
     and 0.05 (dashed line {\it - $D3$}). The grey vertical dashed
     line indicates when the driving velocity is switched off in all
     experiments except {\it $D4$}.}
\end{figure}
The evolution of the total magnetic and kinetic energy is shown in
Fig. \ref{mag_driv.fig} for each experiment $D1$-$D5$. In
Fig. \ref{mag_driv.fig}a, the magnetic energy relative to the
potential energy, as a function of scaled time, is plotted; this is
effectively the `free' magnetic energy in the system. Clearly, this
energy increases in time throughout the driving period of the
experiments with the rate of energy growth increasing with increasing
driving velocity.  Interestingly, the free magnetic energy shows no
significant features that indicate the onset of separator reconnection
(closing the field) or separatrix-surface reconnection (re-opening the
field). This suggests that the injection of magnetic energy through
driving the field is more dominant than the dissipation of the
magnetic energy through reconnection.

Unsurprisingly, the kinetic energy also scales with driving velocity,
however, here there are pronounced differences between the experiments
(Fig. \ref{mag_driv.fig}b). Unlike the magnetic energy, the kinetic
energy initially rises before levelling off and then varies, with only
small fluctuations, until the end of the driving. The time of levelling
off gets later, by a factor of two, as the driving speed doubles,
indicating that its onset is related to the travel time of information
through the domain (the times of levelling represent the same absolute
time in non-scaled time units).  Note, that in the higher
resolution experiment $D5$ the kinetic energy levels off at a higher
value than in $D2$. This is due to the higher velocities that can be
achieved following the formation of narrower current sheets in the high
resolution experiment.

By comparing the two graphs in Fig. \ref{mag_driv.fig} it is clear
that the maximum total kinetic energy in any one experiment is an
order of magnitude smaller than the corresponding magnetic
energy. This implies that the kinetic energy does not contribute
significantly to the heating of the plasma through viscous
dissipation. Instead, the near constant kinetic energy for the latter
part of the driving period indicates that there is an equipartition
between the energy input from magnetic forces and that lost through
viscous dissipation.

Integrating Eq. (\ref{pf.eq}) from $t=0$ to $t=t_d$ suggests that all
experiments should be injected with the same energy under the
assumptions assumed to derive Eq. (\ref{pf.eq}), but this is clearly
not true. There are three reasons why this is so.  First, as seen in
Fig. \ref{pf_driv.fig}, the energy input increases with driving
velocity due, basically, to the differences in the Poynting fluxes at
the onset of driving.  Second, the faster the driver, the shorter the
time available for dissipation and the transfer of energy from
magnetic to kinetic.  Third, the more quickly driven experiments have
a greater angle $\phi$ between the field lines and the surface normal
as information has less time to propagate into the domain resulting in
a slower dissipation rate. Thus, the more quickly driven experiments
have a some what larger build up in magnetic energy.

The curves that continuously rise in both Fig. \ref{mag_driv.fig}
graphs are, of course, related to $D4$ and show that continuous
driving leads to continual energy injection into the domain. This
implies that there is still a long way to go before a time averaged
balance is reached between the Poynting flux, the dissipation and the
conversion of magnetic energy into kinetic energy. In $D4$ the
magnetic energy continues to increase smoothly until at least t=120
when the experiment is stopped.  The kinetic energy also increases
throughout this time, but in a stepwise fashion. This is due to the
fact that the sources are driven many times across the box resulting
in the closing and opening of the fluxes multiple times during the run
with each time more kinetic energy being built up in the system.

Increasing the numerical resolution increases the `free' magnetic
energy in the system, because of a decrease in the joule dissipation
following the associated decrease in magnetic resistivity (seen in the
following section).

To get an idea about the decay times for the various experiments, no
scalings are applied to the time axes for $t>t_d$.  The decay curves
for the magnetic energy show that the more free energy there is
available, the faster it decays. They also indicate that the
experiments all relax to quasi-static states that have approximately
the same energy --- just over 107\% of the initial magnetic energy in
the system.  This implies that (i) the rate of magnetic
dissipation tends to zero for large times and that (ii) a new
force-free topology is found which is clearly different from both the
initial potential configuration and the equivalent configuration found
after the potential evolution of the field. This higher energy level
state is a consequence of the increase in magnetic helicity in the
numerical experiments during the advection of the sources; a helical
structure is clearly visible within the closed field region of each of
the experiments. Such a twist can not be found in the comparable
potential solutions of the time dependent problem. Therefore, since
dissipation of helicity, in general, takes place on a longer time
scale than magnetic dissipation (\cite{Berger84}) the energy of the
magnetic field after the driving has stopped is naturally higher than
before the driving was initiated.

The decay in the kinetic energy behaves in a similar manner to that of
the magnetic energy in that, after the drivers are switched off, the
fall off in kinetic energy is greatest for the experiment with the
largest kinetic energy. Here, of course, it is expected that the
kinetic energy will eventually drop to zero, the level it started at.

\subsubsection{Joule versus Viscous Dissipation}
\begin{figure}
\epsfxsize=7.cm
\epsfbox{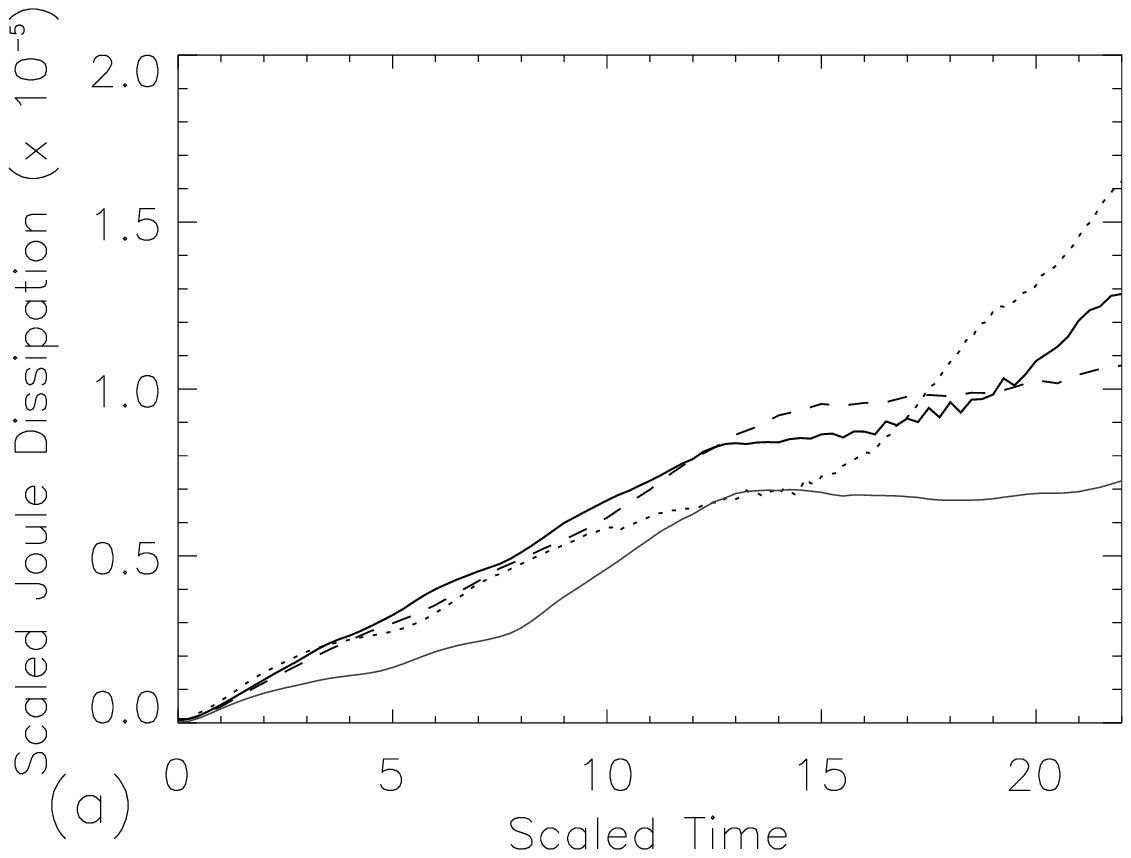}
\epsfxsize=7.cm
\epsfbox{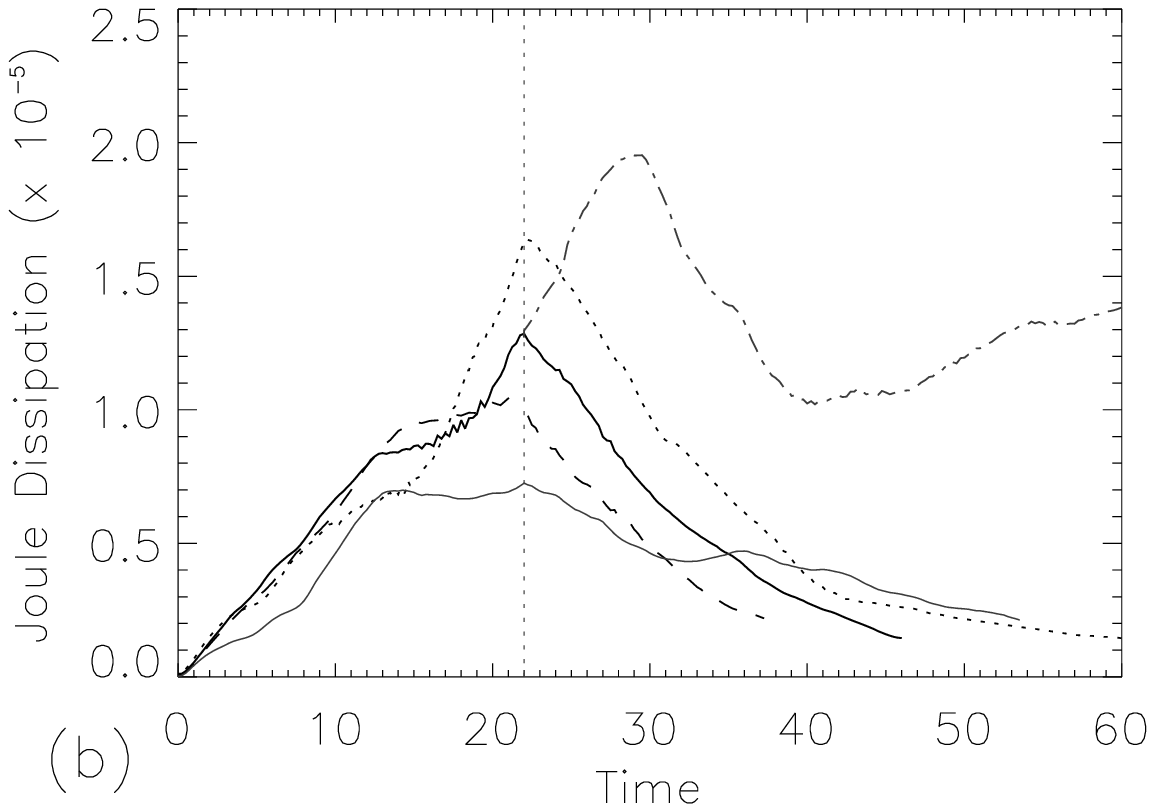}
\epsfxsize=7.cm
\epsfbox{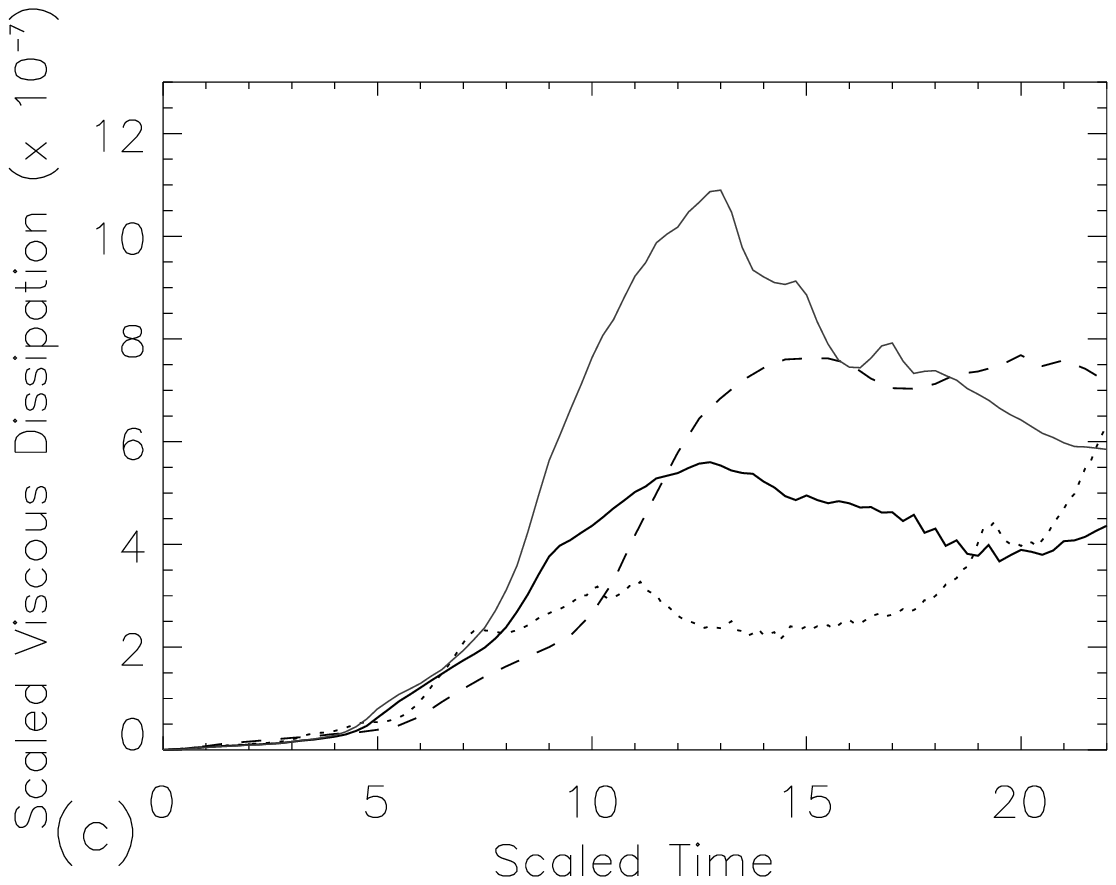}
\epsfxsize=7.cm
\epsfbox{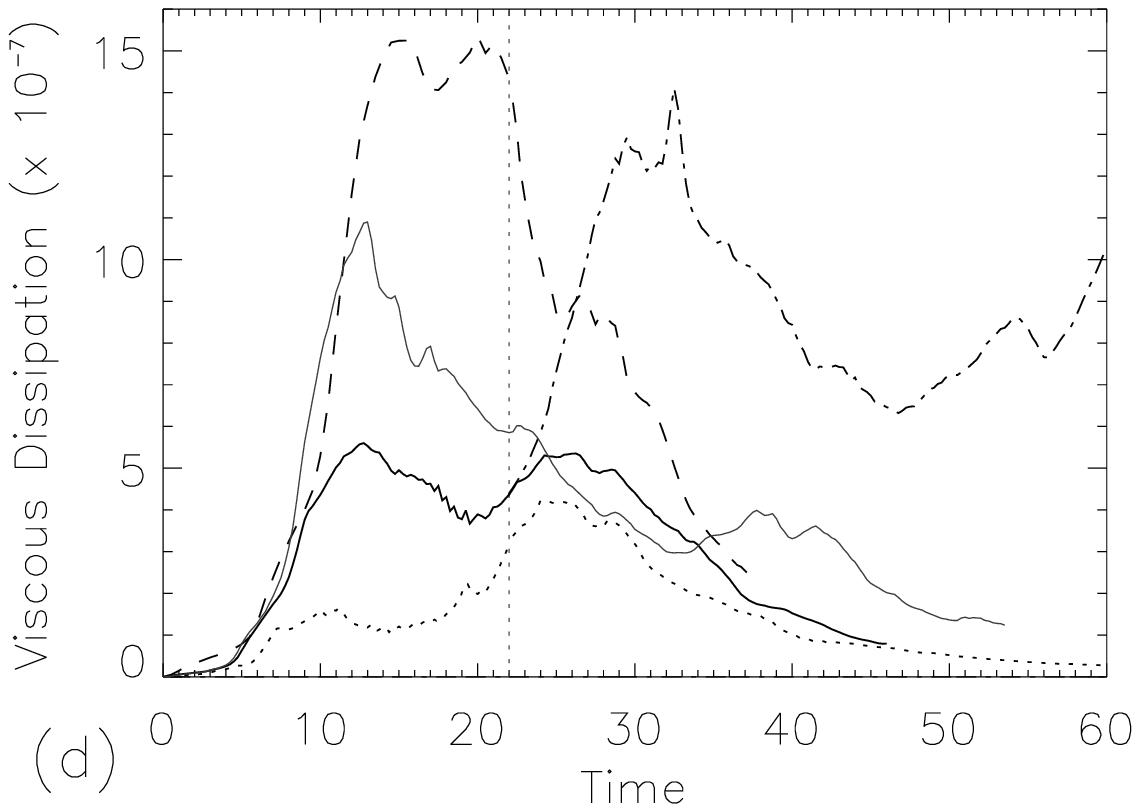}
\caption[]{\label{joule_driv.fig} (a) and (c) The scaled Joule and
viscous dissipation, respectively, as a function of scaled time for
$D1$-$D5$ up until the driver is stopped at $t=t_d$. (b) and (d) The
correct Joule and viscous dissipation, respectively, but with the time
scaled for $t<t_d$ and with correct time scales after $t=t_d$.  The
velocities are 0.0125 (dotted line {\it $D1$}), 0.025 (thick solid
line {\it $D2$}, thin solid line {\it $D5$} and dot-dashed line {\it
$D4$}) and 0.05 (dashed line {\it $D3$}.)}
\end{figure} 
As already suggested, joule dissipation is likely to dominate viscous
dissipation. Fig.  \ref{joule_driv.fig}, which shows the scaled joule
and viscous dissipation up until the drivers are switched off in the
two left-hand graphs, confirms this with the dissipation due to ohmic
heating some 50-100 times more effective that that due to viscous
effects.  Fig. \ref{joule_driv.fig}a shows that three of the
experiments, $D1$-$D3$, have almost identical scaled joule dissipation
rates that increase linearly up to t=10. At $t=10$, $D1$, the most
slowly driven case, then diverges from this linear regime with the
other two cases following suit with delays of about two scaled-time
units per increase in driving velocity. The times of these changes are
comparable to the time at which about half of the flux between the two
sources becomes connected (see Section (\ref{drive_connect.sec}) for a
detailed discussion of the changes in connectivity) and the current in
the separator current sheet starts to slowly decrease.  These three
experiments then have a period of fairly constant dissipation, lasting
about 4-5 scaled-time units, before once again increasing. This second
increase in dissipation relates to the re-opening of the magnetic flux
through separatrix-surface reconnection (again see Section
(\ref{drive_connect.sec}) for further details).

The viscous dissipation curves (Fig. \ref{joule_driv.fig}c) also all
start off with a similar linear growth rate, but they diverge from
this after approximately one Alfv{\'e}n time. Each curve then peaks at
about the time the joule dissipation curves first break from their
initial linear regime. Thus, these peaks are related to the closing of
the field which results in highly localised velocity structures from
the strongly driven separator reconnection. After the peaking the
viscous dissipation rates either level off or drop, although their
behaviour is not completely clear. There is no obvious counter part to
the re-opening process, which is observed in the joule dissipation
curves as a second rise phase. This is rather surprising and is likely
to be related to the fact that the re-opening process occurs through
weakly driven separatrix-surface reconnection which produces diffuse
velocity structures.

Integrating the total joule dissipation from the start to the end of
the driving gives close to the same value for experiments $D1-D3$.
This indicates that approximately the same amount of energy is
available for heating the plasma independent of the speed at which the
sources are driven.  This is not that surprising since in each case
one would expect the same amount of flux to close and re-open and it
relates well to the first-order assumption of the integrated energy
input being independent of the rate of driving.  Thus the heating
essentially depends on the duration of the energy release process. If
effects such as anisotropic heat conduction and optically thin
radiation are ignored then the plasma heating rate would, to a first
approximation, scale inversely with the duration time of the event.

The right-hand graphs in Fig. \ref{joule_driv.fig} show the unscaled
joule and viscous dissipation rates against scaled time for $t<t_d$
and unscaled time for $t>t_d$.  The joule dissipation curves
(Fig. \ref{joule_driv.fig}b) reveal that the most rapidly driven case
naturally has a much higher rate of dissipation than the slower
cases. Indeed, there is roughly a factor of 2 increase in rate for
each doubling in driving speed.  After the driving is stopped the
joule dissipation rapidly decreases with the decay rate scaling with
the driving speed. It is clear that the rates of joule dissipation in
the $D1$-$D3$ experiments all decay to zero after about the same
time. This is consistent with the evolution of the magnetic energy
discussed earlier which approaches a new constant level towards the
end of the experiments. The time scale for this to occur is short
relative to large-scale magnetic diffusion. From the traditional
diffusion equation, one might expect an exponential decay of the
dissipation, but the decay curves are more reminiscent of a series of
near linear decay periods changing at regular intervals of
approximately twice the crossing time for the domain.
The rapid decay of the dissipation is due to the diffusivity used in
the numerical code, which has contributions from both a 2nd and 4th
order diffusion operator implemented to stabilise the high-order
finite-difference scheme. The high-order diffusivity acts efficiently
on length scales close to the resolution limit, whilst length scales
much longer than the resolution limit feel very little
dissipation. This allows for the possibility that large-scale current
structures can maintain their strength almost unaffected by
dissipation.

The curves in Fig. \ref{joule_driv.fig}d are the viscous dissipation
counter parts of those in Fig.  \ref{joule_driv.fig}b. and they behave
in a similar fashion with experiment $D3$ naturally having a much
higher viscous dissipation rate than the other cases. All the viscous
dissipation rates tend to zero at the end of the experiments with the
dissipation in $D3$ falling off fastest. The decay of the viscous
dissipation rates is consistent with the very low kinetic energies
seen at the end of all the experiments.

Once again experiment $D4$ is clearly visible as having the only
curves in Figs \ref{joule_driv.fig}b and \ref{joule_driv.fig}d that
continue to increase after the driving has stopped for the
other experiments. These curves show
both that joule and viscous dissipation is maintained, despite
significant fluctuations, whilst driving continues. It is most likely
that the fluctuations are related to the changes in reconnection that
occur due to the crossing of the periodic boundaries, although this is
not easy to follow.

\subsubsection{Peak Flow Velocities}
\label{drive_conenct.sec}
\begin{figure}
\centering \epsfxsize=6.5cm
\epsfbox{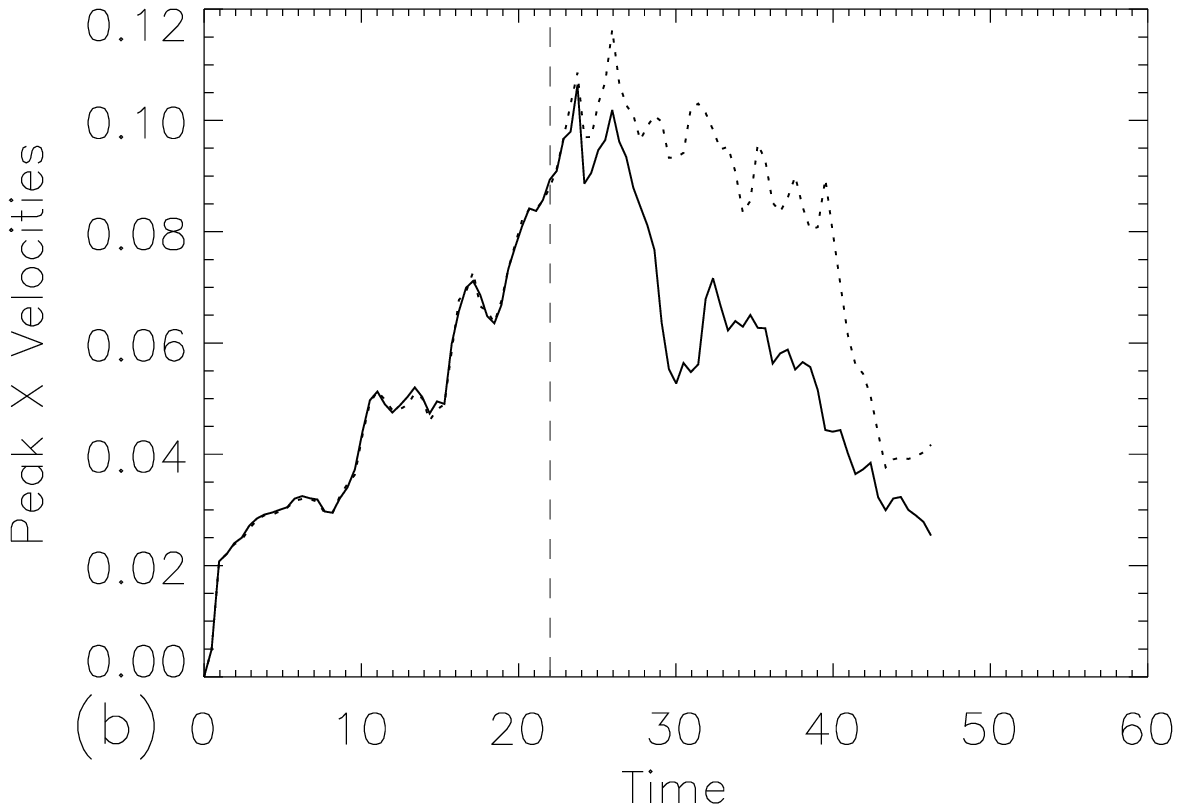}
\epsfxsize=6.5cm
\epsfbox{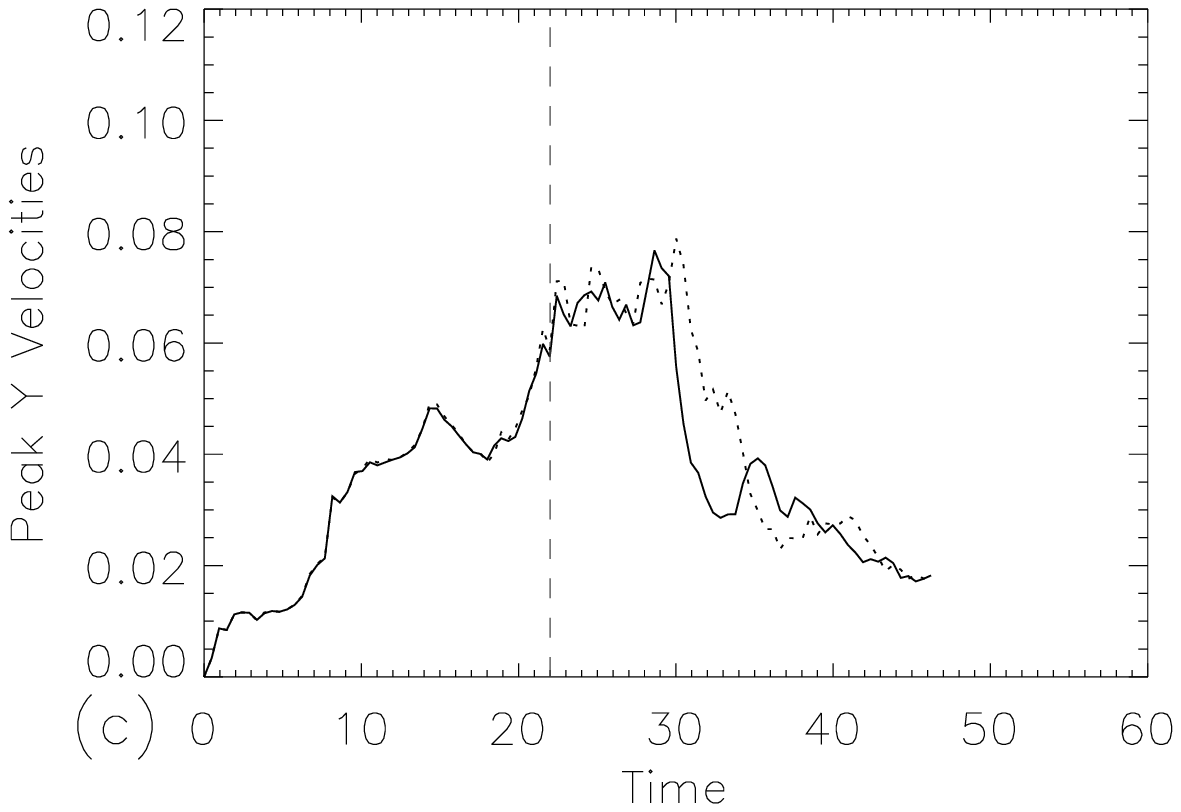}
\epsfxsize=6.5cm
\epsfbox{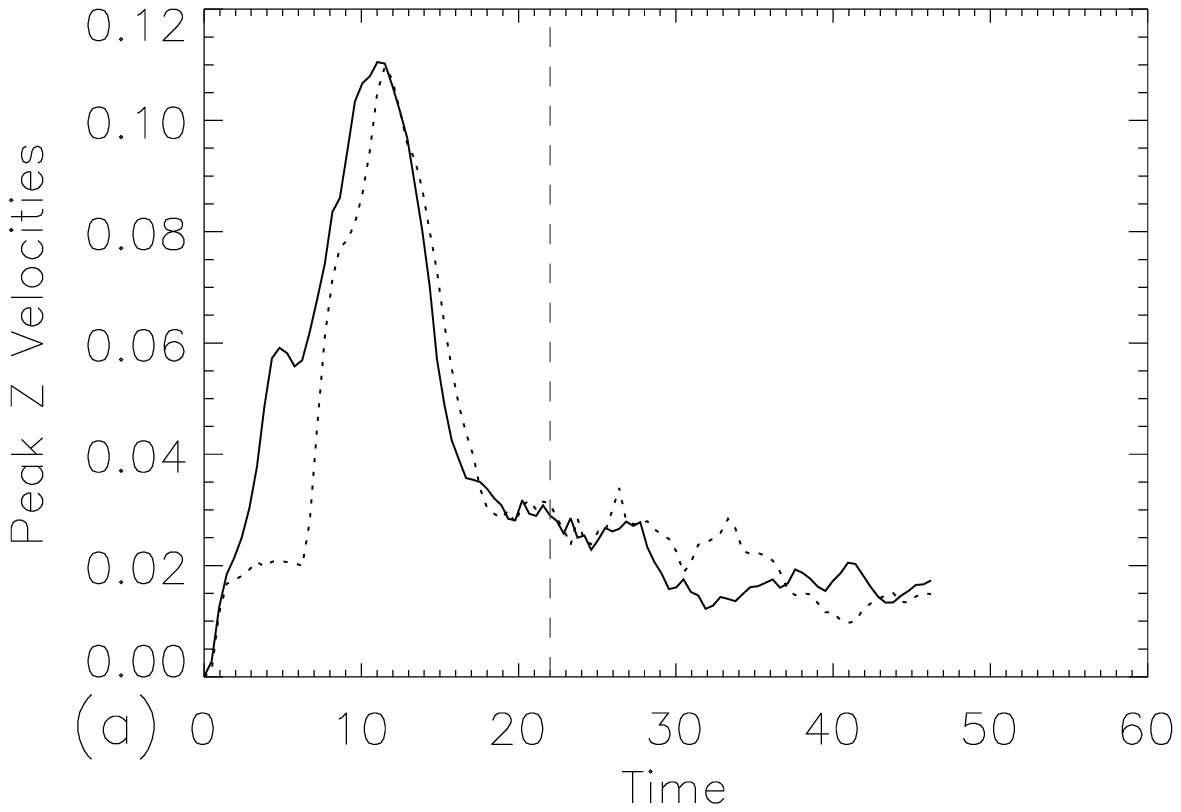}
\caption[]{\label{drive_1_vel.fig} The time evolution of the absolute
  values of the maximum (solid) and minimum (dotted) velocities in the
  (a) $x$, (b) $y$ and (c) $z$ directions of the $y=0$-plane half way
  between the driving boundaries for experiment $D2$ which has a driving
  velocity of $0.025$.}
\end{figure}
Finally, we investigate the temporal variations in the maximum and
minimum flow velocities of the $y=0$-plane half way between the
sources (i.e., in the vertical plane perpendicular to the direction of
driving). The evolution of all the experiments follows the same
pattern, thus we limit our discussion to just experiment $D2$.
Fig. \ref{drive_1_vel.fig} shows the extremes of the three velocity
components as functions of time.  Three significant features are
noted. Firstly, the $z$ velocity (Fig. \ref{drive_1_vel.fig}c) shows a
clear broad peak in maximum and minimum velocity followed by a period
at a constant low level.  Secondly, both the $x$ and $y$ velocities
(Fig. \ref{drive_1_vel.fig}a,b) show an even broader peak in their
maximum/minimum values starting after the peak in the $z$
component. Thirdly, all components eventually level out to, or
decrease to, a low, almost identical, constant velocity after a period
of time. Comparing the evolution of these maximum/minimum velocity
curves with the temporal evolution of the field line connectivity
reveals clear correlations.  The peaks in the $z$ velocity components
relate to the initial closing of flux in which the two flux lobes
start to interact. Thus, the peak $z$ velocities are most likely the
outflows from the separator reconnection suggesting that there are two
oppositely directed jets with almost the same maximum velocity. These
reconnection jets cease to be important contributers to the plasma
dynamics at about the time when the closed flux peaks indicating that
there is a strong decline in the closing process after this. This
feature is clearly evident in all experiments, independent of the
driving speed.

The structure of the $x$ and $y$ velocity components vary much more
between the experiments.  These velocities are related to the
re-opening of the closed magnetic field and represent the mainly
horizontal outflow velocities associated with this process.  Finally,
a low level velocity state is reached for all the velocity components
once the process of re-opening the flux ceases through lack of
driving.

From the point of view of scalings it is found that the absolute peak
$z$ velocity, $v_{zmx}$, increases with increasing driving velocity at
a rate that is slower than the square root of the driving velocity
(see Table (\ref{peak_v.tab})).  However, the exact nature of this
scaling relation is not known.
\begin{table}[h]
\caption{\label{peak_v.tab} The absolute peak $z$ velocity, $v_{zmx}$,
for the four experiments}
\begin{tabular}{c l l l l} \hline
                    & $D1$   & $D2$  & $D3$ & $D5$ \\ \hline 
	    $v_{d}$ & 0.0125 & 0.025 & 0.05 & 0.025 \\ 
	    $v_{zmx}$ & 0.067  & 0.11  & 0.15 & 0.16 \\ \hline
\end{tabular}
\end{table}

\subsection{Effects of the Strength of the Overlying Magnetic Field}
In this section, we investigate the effects on the energy release of
having different strengths of overlying magnetic field in the same
magnetically interacting system as above.  Changing the field strength
is equivalent to varying the flux from the sources and, hence, to
varying the size of the flux lobes.  The stronger the overlying field
the smaller the flux lobes and the more rigid their advection.
Clearly, this results in a smaller region of the system being directly
influenced by the advection of the sources.  Predictions of the
behaviour of the various physical quantities are, therefore, not
straight forward and we conduct the following numerical experiments to
try and determine the possible scaling relations.

Four experiments, labelled $H1$-$H4$, are carried out with the first
three ($H1$-$H3$) having overlying field strengths of $B_y=0.076$,
$B_y=0.1$ and $B_y=0.2$, respectively. Experiment $H4$ is the same as
$H3$ except it has double the numerical resolution. In all cases, the
driving velocity and peak source strengths are the same (See Table
\ref{cases.tab} for details). Note, also that experiments $D2$ and
$H2$ are the same. Here, since the driving velocity is the same in
each experiment, there is no need to scale any of the results to
different time frames, so no scalings are applied.

\subsubsection{Poynting Flux}
\begin{figure}
\centering
\epsfxsize=7.cm \epsfbox{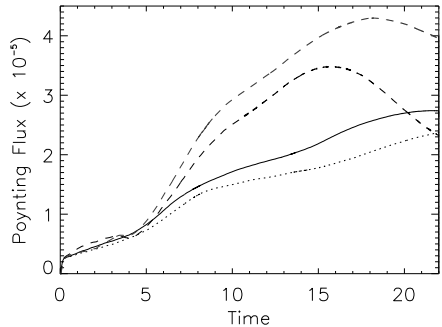}
\caption[]{\label{pf_height.fig} Poynting flux variations in time for
  experiments $H1$-$H4$. The different field strengths are $B_y=0.076$
  ($H1$ - dotted), $B_y=0.1$ ($H2$ - solid black) and $B_y=0.2$
  ($H3$ - thick dashed and $H4$ - thin dashed).}
\end{figure}
Varying the strength of the overlying field clearly has an affect on
the Poynting flux (Fig.  \ref{pf_height.fig}) which is anticipated
since, from Eq. (\ref{B_induction.eq}), it is apparent that increasing
the strength of $B_y$ can change the growth rate of $B_x$, and this in
turn effects the Poynting flux (Eq. (\ref{Pf_plane.eq})). This effect,
however, only kicks in after one Alfv{\'e}n crossing time
(approximately, 5 time units), implying that $u_x$ has only a small
variation in the $y$-direction at the driving boundary - this is not a
surprise considering the imposed driving profile.

After t=5, the Poynting flux increases. This increase in Poynting flux 
is triggered by the
crossing of the domain by the wave front created by the initial ramp
up of the driving velocity. The stronger the overlying field, the
smaller the source flux lobes, and although the region affected
by the wave fronts is smaller their effect is greater. Thus, the
effect on the local field lines from the wave interaction increases
with increasing $B_y$. Hence, experiment $H3$, which has the strongest
overlying field, has the fastest growth in Poynting flux. Furthermore,
$H4$, which has a higher numerical resolution and therefore less
numerical diffusion than $H3$, has a larger Poynting flux than $H3$,
because it produces greater changes in the field line orientation.

For experiments $H1$ and $H2$, the increase in Poynting flux continues
until the driving is stopped at t=22, but for $H3$ and $H4$, the
growth in Poynting flux peaks at t=16 and t=18, respectively. The
decline seen in $H3$ and $H4$ occurs because these experiments show a
faster restructuring of the magnetic field to a lower energy state
(more rapid release in magnetic energy) than the two other experiments
giving rise to the early decay in Poynting flux.

This implies that making simple predictions about any scaling is
difficult as it depends on the detailed structures of the magnetic
field topology and velocity driving profile at any one
time. Determining the overall magnetic topology is relatively straight
forward, but the detailed small-scale variations in the magnetic field
are not so easy to understand and yet they may be more
important. Similarly, mean flow patterns created by the driving
profile may be determined, but with addition of local variations means
large errors may arise in estimates of the Poynting flux for real
situations.

\subsubsection{Magnetic and Kinetic Energy}
\begin{figure}
\epsfxsize=7.cm
\epsfbox{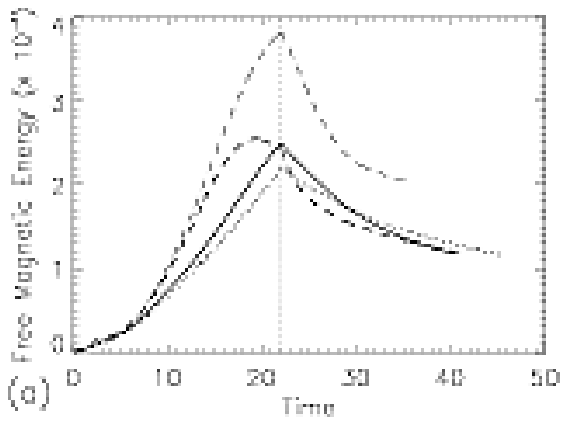}
\epsfxsize=7.cm 
\epsfbox{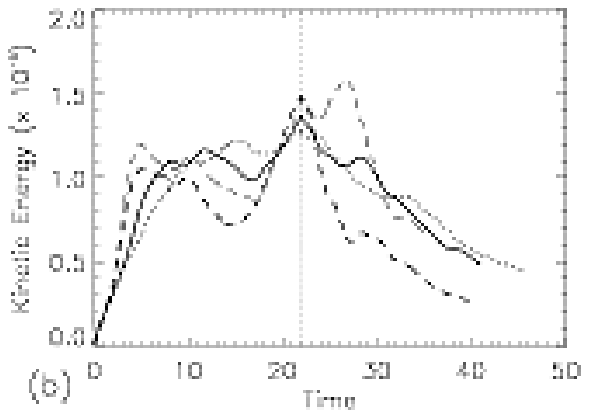}
\caption[]{\label{mag_height.fig} (a) The 'free' magnetic energy and
   (b) the kinetic energy for experiments $H1$-$H4$. The field
   strengths are $B_y$=0.076 ($H1$ - dotted), $B_y$=0.1 ($H2$ solid)
   and $B_y$=0.2 ($H3$ - thick dashed and $H4$ - thin dashed). The
   vertical dotted line denotes when the driver is switched off.}
\end{figure}
Naturally, increasing the strength of the overlying field increases
the total magnetic energy in the system. It also increases the amount
of `free' magnetic energy (i.e., magnetic energy minus the potential
magnetic energy) available at any time
(Fig. \ref{mag_height.fig}a). The difference in `free' magnetic energy
between all the experiments only becomes apparent after one Alfv{\'e}n
crossing time as this is when the Poynting flux curves start to
diverge. $H3$ and $H4$ clearly have a much larger gain in magnetic
energy than the other experiments, as one might expect from their
increased injection of Poynting flux. Note, though, that the `free'
magnetic energy for $H3$ starts to decrease at t=19, which is shortly
after the time at which the Poynting flux starts to decrease. In
comparison, the `free' magnetic energy of the high resolution run,
$H4$, continues to increase until the driving is turned off and does
not follow the early decline that its Poynting flux has.

After the driving has stopped, the magnetic energy decreases for all
experiments, with the most rapid decline seen in $H3$.  The reason for
the different rates of magnetic energy decline from experiments $H1$
to $H3$ is, in part, related to the relative decrease in numerical
resolution of the flux lobes as the overlying field strength
increases. $H3$ shows stronger local currents coupled with a smaller
spatial resolution leading to a faster numerical dissipation of the
currents in the system and, therefore, a more rapid decrease in the
`free' magnetic energy.  This is confirmed by comparing the results
from $H3$ and $H4$. The decay rates in these two experiments look
similar, but the high resolution experiment is capable of maintaining
a higher `free' magnetic energy than the low resolution experiment.

For the varying driving velocity experiments the magnetic field
ultimately appears to approach a steady-state with a higher magnetic
energy than the initial state. Such a phase is not reached here, as
the experiments are not followed for quite as long. However, it is
anticipated that these experiments would also reach similar near
steady states, but that the magnetic energies of these states in each
experiment would be different, due to their different initial magnetic
energies. In deed, since $D2$ and $H2$ are the same experiment we know
$H2$ drops to a steady state.

The temporal evolution of the kinetic energy
(Fig. \ref{mag_height.fig}b) follows basically the same pattern in
each of the four experiments here. Despite the different initial
growth rates in kinetic energy the experiments all level out at
approximately the same energy and fluctuate around this. This, as
explained in the driving velocity experiments, is related to the
process of separator reconnection (closing the field). All the
experiments then dip before once again peaking at a higher energy
level at around the time the driver is switched off. This second rise
in kinetic energy is related to the separatrix-surface reconnection
process (re-opening the field); it is cut short (as seen in
Fig. \ref{rec_driv_height.fig}) in a number of the experiments due to
the switching off of the driver. 

As for the magnetic energy, the kinetic energy decreases rapidly after
the driving ceases with the most rapid decay seen in experiment
$H3$. This is again most likely to be related to the effectively lower
numerical resolution in the region of interest in $H3$. This
interpretation is supported by comparing
$H3$ and $H4$, where the latter has a second peak at t=27 after which
is follows the same trend as $H3$, just delayed about 5 time units.

Also, the kinetic energy in these experiments is a factor of 10
smaller than their `free' magnetic energy, as it is in the varying
driving velocity experiments. Thus, not surprisingly, the viscous
dissipation in these experiments is weak. Furthermore, since the
evolution of the kinetic energy is similar in all the experiments it
is not surprising that the peak flow velocities for each experiment
are also similar and, hence, no scalings with overlying field strength
of these peak velocities are found. As in the varying driving velocity
experiments strong vertical reconnection jets are observed followed by
slightly weaker horizontal jets corresponding to the separator
reconnection and separatrix-surface reconnection, respectively.

\subsubsection{Joule versus Viscous Dissipation}
\begin{figure}
\epsfxsize=7.cm
\epsfbox{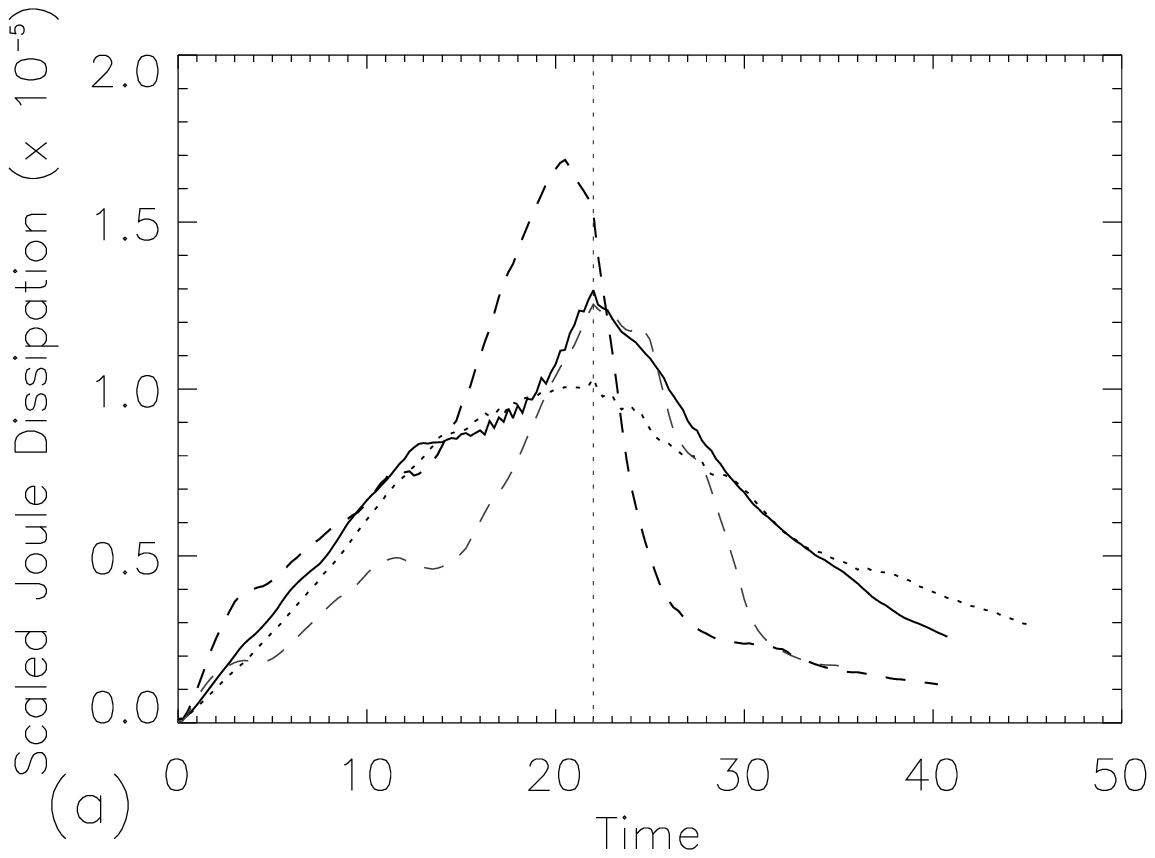}
\epsfxsize=7.cm
\epsfbox{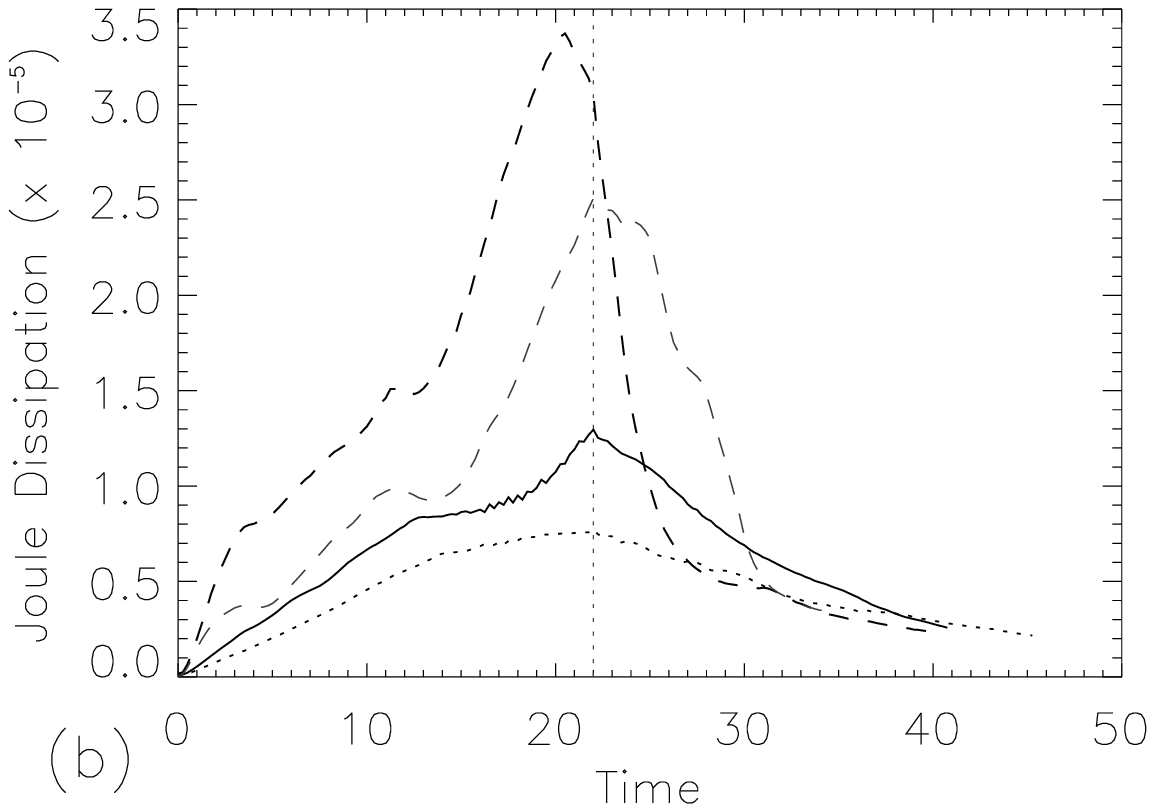}
\caption[]{\label{joule_height.fig} The joule dissipation for the
   $H1$-$H4$ experiments.  Left the dissipation scaled with the
   strength of the $B_y$ component. Right the unscaled data.  The
   lines represent the different field strengths: $H1$, $B_y$=0.076
   dashed, $H2$, $B_y$=0.1 dotted and $H3$ and $H4$ $B_y$=0.2 thick 
   dashed and thin dashed.}
\end{figure}
As already mentioned the viscous dissipation is considerably weaker
than joule dissipation. Here, therefore we basically confine our
discussion to the joule dissipation which varies with varying
overlying speed strength.  In Fig. \ref{joule_height.fig}, the
left-hand graph shows the Joule dissipation scaled by the strength of
the initial $B_y$ field, whereas in the right-hand graph the real
values of the dissipation are plotted for the full duration of the
experiments.  Up until $t=15$ the three scaled dissipation curves,
$H1$-$H3$, in the left-hand graph all appear to be very similar
indicating that the differences between the joule dissipation in the
different experiments simply result from a linear scaling in the $y$
component of the magnetic field. The curve for experiment $H4$ on the
other hand shows, as expected, that the magnetic dissipation decreases
as the numerical resolution increases. This leads to a slight delay in
changes in the dissipation rate of the high resolution case with
respect to the low resolution one.

What is also clear from the above discussion and the right-hand panel
of Fig. \ref{joule_height.fig} is that the more concentrated the
interacting magnetic flux lobes, the greater the dissipation per unit
time. This is not altogether surprising since the Lorentz force
squeezing the two flux regions together increases with increasing
$B_y$. Therefore, the strength of the current sheet also increases
and, hence, the current reaches the numerical limit earlier in the
larger $B_y$ case thus initiating reconnection sooner.

After t=15 the dissipation in $H3$ and $H4$ increases
significantly. This is approximately the time that the Poynting flux
and magnetic energy start to decrease and the kinetic energy
increases, indicating as previously seen by other measures, that this
is a period of enhanced dynamical activity involving magnetic field
restructuring (i.e., part of the separator-reconnection phase).

After the driver is switched off the dissipation rates fall
(right-hand graph in Fig. \ref{joule_height.fig}), with the most rapid
decline seen in $H3$ and $H4$ such that all experiments, with the same
numerical resolution, appear to tend towards the same joule
dissipation rate. Experiment $H4$, which has a higher numerical
resolution than the other three experiments, is not run for long
enough to determine the eventual lower dissipation rate. This trend
towards similar dissipation rates at the end of the experiment is
similar to the behaviour seen in the driving velocity experiments and
is just a sign that eventually all experiments tend towards
quasi-static states without concentrated current systems, and hence,
very little dissipation.

Comparisons with the viscous dissipation rates for these experiments
show that the joule dissipation rates are on the order of 20 times
larger than the viscous dissipation rates. Joule dissipation is,
therefore, globally dominant, although it is possible that at some
localised points (usually where both viscous and joule dissipation are
weak) the reverse may be true.

\section{Conclusion} 

From these experiments of driven reconnection between two
initially unconnected flux sources we find a number of important
results. First, from the experiments with varying driving velocity we
find that:
\begin{itemize}
\item To first order, the energy imposed on the system is
independent of the driving velocity when starting from a simple
potential configuration. That is, the amount of energy injected
is proportional to the distance travelled, not the speed at which the
distance is covered.
\item The `free' magnetic energy available for release is governed
mainly by the distance travelled rather than the speed of travel,
however, the faster driven experiments do have marginally more `free'
energy than the slower driven ones. This means that per unit time the
`free' magnetic energy accumulated increases with increasing driving
velocity, though this increase is at a slower rate than the increase
in velocity itself.
\item Similarly, the rate of Joule dissipation is related to the
distance travelled implying that rapidly driven foot points lead to
bright, intense, but short-lived events, whilst slowly driven foot
points produce weaker, but longer-lived brightenings. Integrated over
the lifetime of the events both would produce approximately the same
heating if all other factors were the same. 
\item The deviations from the potential evolution of the magnetic
field increase with increasing driving velocity. This is as one might
expect since in the more slowly driven experiments the magnetic field
has more time to reconfigure itself to a more relaxed state as the
driving persists.
\end{itemize}

From the experiments analysing the effect of varying the magnitude of
the overlying magnetic field the following results are found:
\begin{itemize}
\item Initially the Poynting flux is independent of the strength of
the overlying magnetic field strength. 
\item The `free' magnetic energy increases with increasing strength of
the overlying magnetic field, as expected, since the pointing flux,
after a while, increases with increasing overlying field strength.
\item The reconnection rate of the experiments increases with
increasing overlying field strength, because of the greater
confinement and higher field strengths of the flux lobes leading to a
faster Alfv{\'e}n speed in the system. Exactly how effective this is
is unclear since at the same time the resolution across the lobes
reduces as they become more compact, effectively increasing the
resistivity (a numerical artifact). Any dependence of the reconnection
rates on the strength of the overlying field is likely to be weak.
\item The rate of reconnection decreases with increasing numerical
resolution and decreasing magnetic resistivity.
\end{itemize}

Observations that relate to both experiments are:
\begin{itemize}
\item Non-linear effects and deviations from the simple potential
evolution make it increasingly difficult to provide precise
predictions of the energy input and energy release rates.
\item After the driving is stopped, the system relaxes toward an
energy state with a higher energy than the initial, or even the
equivalent, potential magnetic configuration.
\item There is a correlation between peak velocities found in
particular directions and the two classes of magnetic reconnection
that occur. Thus, indicating that, not surprisingly, the initial
separator reconnection (closing of the field) produces stronger
velocities than the subsequent separatrix-surface reconnection
(opening of the field).
\item The numerical resolution of the experiments influences the
dynamics and evolution of the system, but only marginally.  
The higher resolution experiments have a lower magnetic resistivity
and the reconnection rate becomes slightly slower than in the
comparable low resolution experiments. This indicates a weak
scaling of the reconnection rate with the magnetic Reynolds number.
We are not at present capable of following this avenue numerically to
determine in more detail how the scaling depends on the
resistivity, $\eta$.
\item The effect of changing the numerical resolution with a
factor of two, and through this the effective value of $\eta$, has a
different impact on the two reconnection scenarios. In the
strongly-forced closing phase, where separator reconnection occurs,
the difference in reconnection rate is small.  While, for the
less-stressed reopening, due to separatrix-surface reconnection, the
effect is much more pronounced.  Strongly-forced reconnection events
appear therefore to have a weaker dependence on $\eta$ than
less-stressed diffusion events.
\item The experiments show that it is not enough to know the positions
and flux distribution of photospheric sources to predict the
evolutions of the magnetic field. The direction and field strength of
the coronal magnetic field plays a very important role, and only when
this is taken into account may realistic results be obtained.
\item The slower the imposed driving velocity relative to the
Alfv{\'e}n velocity the more `complete' the reconnection
process, i.e., the higher the peak fraction of closed flux or the more
flux from the source involved in reconnection. Therefore, this ratio
(which is affected by $v_d$ and $B_y$) plays a crucial role in
determining how complicated the magnetic field line connectivity can
become.
\end{itemize}

These points show that energy considerations using potential models
can provide an insight into the amount of energy available through
reconnection, but do not provide information on the duration or onset
of the process. To determine these more realistic models of the events
are required. Indeed, the location and time duration of the energy
release and the corresponding effects on the local plasma parameters
are strongly dependent on the non-linear evolution of the magnetic
field.  The experiments also strongly suggest that as the magnetic
field starts to deviate from a simple potential state, then the
possibility of providing realistic predictions using potential models
diminishes and the effects of the full nonlinear evolution
have to be taken into account for making detailed estimates.

\section{Future Perspectives}
\begin{figure}
 \epsfxsize=6.7cm \epsfbox{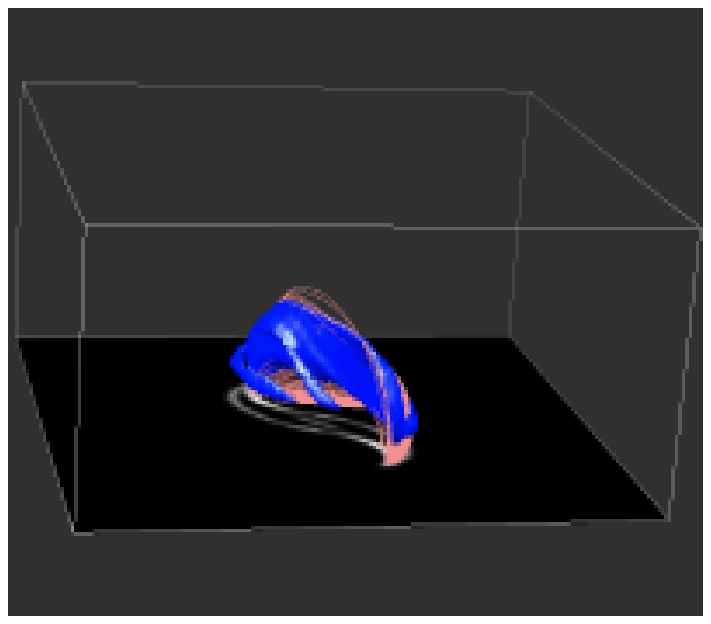}
 \epsfxsize=6.7cm \epsfbox{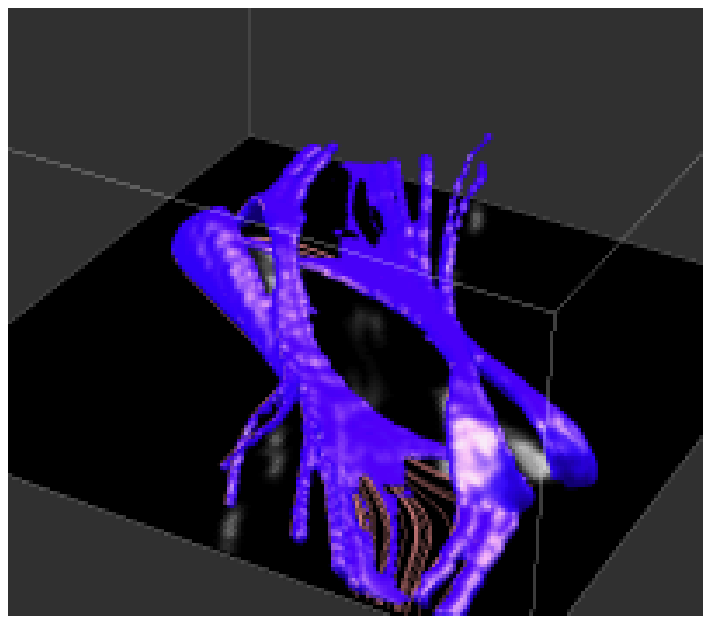}
 \caption[]{\label{temp_maps.fig} The two panels show the structure of
   the heated plasma at different times during the evolution of the
   flux systems in one experiment.  Only the regions with temperatures
   above the initial coronal temperature are shown.  Left, closing of
   flux. Right, opening of flux.  \cite{Galsgaard_Parnell_soho15}
 }
\end{figure}
As the energy equation in the experiments is very simple, it is not
possible to make direct predictions as to the observational appearance
of the flux interaction event. Despite this it is possible to make
simple images of the situations and draw some conclusions. From the
connectivity analysis the field lines changing connectivity are
known. Assuming that an amount of energy proportional to the base flux
of the field lines is released in the reconnection event, and
subsequently distributed along these field lines, it is possible to
make simple temperature maps of the corona. Such maps are shown for
two situations in Fig \ref{temp_maps.fig}.  The left panel represents the
separator reconnection phase where the high temperature region is
confined to the closing part of the field lines, defining a very
compact object. In the right panel the situation is shown for the
separatrix-surface reconnection which opens the connected flux. Here,
the structure is much more fragmented in space and changes
significantly in time.

A clear limitation with this approach is that only regions connecting
to the driving boundary can be illustrated as we have no way of
identifying the field lines that are located in the
corona. Therefore to get a more realistic picture of the observational
appearance of the flux interaction, the model must be improved. A more
realistic model atmosphere is required allowing the reconnection to take
place in a transition region or coronal environment. Furthermore, the
effects of optically thin radiation and anisotropic heat conduction
need to be included. These are effects we would like to address in a
future paper.

\begin{acknowledgements}
KG would like to thank the Carlsberg foundation for financial support.
CEP would like to thank PPARC for financial support through her Advanced
Fellowship. The computational analysis for this paper was carried out 
on the SRIF-PPARC funded Beowulf COPSON cluster in St. Andrews.
\end{acknowledgements}

\bibliographystyle{natbib}
\bibliography{aamnem99,kg,ff,aake,phd,parker,reconnection,carpet,loop,waves}

\begin{thebibliography}{}

\bibitem[{Berger }{1984}]{Berger84}
Berger, M., 1984,
\newblock {Geophys. Astrophys. Fluid Dyn.} {30}, 79

\bibitem[{{Dreher } et~al.}{1997}]{Dreher+ea97}
{Dreher}, J., {Birk}, G.~T., and {Neukirch}, T., 1997,
\newblock {A\&A} {323}, 593

\bibitem[{Galsgaard and
  Nordlund }{1996}]{Galsgaard+Nordlund95xc}
Galsgaard, K. and Nordlund, {\AA}., 1996,
\newblock {J. Geophys. Res.} {101}, 13445

\bibitem[{Galsgaard and
  Parnell }{2004}]{Galsgaard_Parnell_soho15}
Galsgaard, K. and Parnell, C., 2004,
\newblock {ESA SP-575: SOHO 15 Coronal Heating}, 351

\bibitem[{Galsgaard et~al. }{2000}]{Galsgaard+ea00carpet}
Galsgaard, K., Parnell, C.~E., and Blaizot, J., 2000,
\newblock {A\&A} {362}, 383

\bibitem[{Goedbloed }{1979}]{Goedbloed79}
Goedbloed, J.~P., 1979,
\newblock {Lecture Notes on Ideal Magnetohydrodynamcis} {Rijnhuizen Rep}, 83

\bibitem[{Goossens and Ruderman }{1995}]{Goossens+Ruderman95}
Goossens, M. and Ruderman, M.~S., 1995,
\newblock {Physica Scripta} {T60}, 171

\bibitem[{Heyvaerts and Priest }{1983}]{Heyvaerts+Priest83}
Heyvaerts, J. and Priest, E.~R., 1983,
\newblock {A\&A} {117}, 220

\bibitem[{Heyvaerts and Priest }{1984}]{Heyvaerts+Priest84}
Heyvaerts, J. and Priest, E.~R., 1984,
\newblock {A\&A} {137}, 63

\bibitem[{Ionson }{1978}]{Ionson78}
Ionson, J.~A., 1978,
\newblock {ApJ} {226}, 650

\bibitem[{Longcope }{1998}]{Longcope98}
Longcope, D., 1998,
\newblock {ApJ} {507}, 443L

\bibitem[{{Longcope} et~al. }{2001}]{Longcope+ea01}
{Longcope}, D.~W., {Kankelborg}, C.~C., {Nelson}, J.~L., and {Pevtsov}, A.~A.,
  2001,
\newblock {ApJ} {553}, 429

\bibitem[{{Mandrini} et~al. }{1996}]{Mandrini+ea96}
{Mandrini}, C.~H., {Demoulin}, P., {van Driel-Gesztelyi}, L., {Schmieder}, B.,
  {Cauzzi}, G., and {Hofmann}, A., 1996,
\newblock {Sol. Phys.} {168}, 115

\bibitem[{Parker }{1957}]{Parker57}
Parker, E.~N., 1957,
\newblock {JGR} pp 509--520

\bibitem[{Parker }{1972}]{Parker72}
Parker, E.~N., 1972,
\newblock {ApJ} {174}, 499

\bibitem[{Parker }{1987}]{Parker87a}
Parker, E.~N., 1987,
\newblock {ApJ} {318}, 876

\bibitem[{Parker }{1988}]{Parker88b}
Parker, E.~N., 1988,
\newblock {ApJ} {330}, 474

\bibitem[{Parnell and Galsgaard }{2004}]{Parnell+Galsgaard04}
Parnell, C. and Galsgaard, K., 2004,
\newblock {A\&A} {428}, 595

\bibitem[{Parnell and Priest }{1995}]{Parnell+Priest95}
Parnell, C. and Priest, E., 1995,
\newblock {Geophys. Astrophys. Fluid Dynamics} {80}, 255

\bibitem[{Parnell et~al. }{1994a}]{Parnell+ea94}
Parnell, C., Priest, E., and Golub, L., 1994a,
\newblock {Solar Phys.} {151}, 57

\bibitem[{Parnell et~al. }{1994b}]{Parnell+ea94b}
Parnell, C., Priest, E., and Titov, V., 1994b,
\newblock {Sol. Phys.} {153}, 217

\bibitem[{Petschek }{1964}]{Petschek64}
Petschek, H.~E., 1964,
\newblock in W. Hess (ed.), {Physics of Solar Flares}, pp 425--439, NASA Spec.
  Publ. SP-50, Washington, DC

\bibitem[{Priest et~al. }{1994}]{Priest+ea94b}
Priest, E., Parnell, C., , and Martin, S., 1994,
\newblock {apj} {427}, 459

\bibitem[{{Shimizu} et~al. }{1994}]{Shimizu+ea94}
{Shimizu}, T., {Tsuneta}, S., {Acton}, L.~W., {Lemen}, J.~R., {Ogawara}, Y.,
  and {Uchida}, Y., 1994,
\newblock {ApJ} {422}, 906

\bibitem[{Sweet }{1958}]{Sweet58}
Sweet, P.~A., 1958,
\newblock in B. Lehnert (ed.), {Electromagnetic Phenomena in Cosmical Physics},
  pp 123--134, Cambridge University Press, London

\bibitem[{van Ballegooijen }{1986}]{vanBall86}
van Ballegooijen, A.~A., 1986,
\newblock {ApJ} {311}, 1001

\end{thebibliography}

\end{document}